\documentclass[aps,twocolumn,floatfix,showpacs,prb,superscriptaddress]{revtex4}

\usepackage{color}
\usepackage{graphicx}
\usepackage{amsmath}
\usepackage{amsfonts}

\begin{document}

\title{Probing Majorana edge states with a flux qubit}

\author{Chang-Yu Hou}
\author{Fabian Hassler}
\author{Anton R. Akhmerov}
\affiliation{Instituut-Lorentz, Universiteit Leiden, P.O. Box 9506, 2300 RA Leiden, The Netherlands}
\author{Johan Nilsson}
\affiliation{Department of Physics, University of Gothenburg, 412 96 Gothenburg,  Sweden}

\date{\today}

\begin{abstract}
A pair of counter-propagating Majorana edge modes appears in chiral
\emph{p}-wave superconductors and in other superconducting systems belonging
to the same universality class. These modes can be described by an Ising
conformal field theory. We show how a superconducting flux qubit attached to
such a system couples to the two chiral edge modes via the disorder field of
the Ising model. Due to this coupling, measuring the back-action of the edge
states on the qubit allows to probe the properties of Majorana edge modes.
\end{abstract}


\maketitle

\section{Introduction}

Chiral Majorana fermion edge states were originally predicted to exist
in the $5/2$ fractional quantum Hall plateau.~\cite{Moore91}
These edge states support not only neutral fermionic excitations but
also more exotic edge vortices. A single edge vortex corresponds
to a $\pi$ phase shift to all fermions situated to one side of it.~\cite{Fendley07,Rosenow08,Rosenow09}
 Two edge vortices may either
fuse into an edge fermion or annihilate each other, with the outcome
depending on the preceding evolution of the system. In other words,
the edge theory (together with the corresponding bulk theory) possesses
non-Abelian statistics.~\cite{Read-Green,Volovik99,Ivanov01,Kitaev03}
This unusual physics and its potential applications to
topological quantum computation are the reasons why the
Majorana edge states have attracted much attention
recently.~\cite{DasSarma05,Stern06,Bonderson06,Kim06,Bena06,Nayak-RMP08}

Similar non-Abelian anyons and their corresponding edge states appear
in superconducting systems as well. Initially it was discovered that
\emph{p}-wave superconductors support non-Abelian anyons in the bulk and chiral Majorana
edge states.~\cite{Read-Green,Volovik03,Kitaev01} Later it was shown that depositing
a conventional $s$-wave superconductor on the surface of a topological
insulator while breaking time-reversal symmetry provides an alternative
route to realize these non-Abelian states.~\cite{Fu08,Fu09,Akhmerov09}
Alternative proposals include substituting the topological insulator
by a two-dimensional electron gas with spin-orbit coupling~\cite{Sau10,Alicea10,Potter10} or
by a half-metal.~\cite{Duckheim,Chung10}
The realizations of Majorana edge states using \emph{s}-wave superconductors have the
following advantages: first, they rely on combining simple,
well-studied ingredients. Second, the materials do not have to be
extremely pure unlike samples needed to support the fractional quantum Hall
edge states. Finally, the superconducting implementations of Majorana
fermions may feature a larger bulk excitation gap and may therefore
be operated at higher temperatures.

The downside of the superconducting implementations of Majorana edge states
is the lack of means to manipulate edge vortices.~\cite{Fu09,Akhmerov09}
Different from the 5/2 fractional quantum Hall state, the edge vortices are not coupled to charge and thus cannot be controlled by applying voltages.~\cite{Nilsson10}
Therefore, the standard proposal to probe the edge vortices in superconducting
systems is to inject fermion excitations into the edge, to let them split
into edge vortices, and finally to conclude about the behavior of the
edge vortices from the detection of the fermion excitations after the
subsequent fusion of edge vortices.\cite{Akhmerov09,Fu09,Nilsson10,Sau10b}

In this paper, we propose a more direct way to manipulate and measure edge
vortices using a flux qubit consisting of a superconducting
ring interrupted by a Josephson junction.~\cite{makhlin:01,Hassler10} Our
main idea is based on the following observations: first, an edge vortex is
created when a superconducting vortex crosses the edge. Second, the motion
of the superconducting vortices can be fully controlled by a flux qubit,
since by applying a flux bias to the qubit one can tune the energy cost
for a vortex being present in the superconducting ring.~\cite{makhlin:01}
In this way, attaching a flux qubit to a system supporting Majorana edge states
allows to directly create, control, and measure edge vortices without
relying on splitting and fusing fermionic excitations.

We note that our proposal is not necessarily advantageous for the
purposes of topological quantum computing since quantum computing
with Majorana fermions may even be realized without ever using edge
states.\cite{Hassler10,Alicea10b,Sau10c} Instead the aim of our investigation
is to develop a better tool for probing the fractional excitations of the
edge theory.

The paper is organized as follows.
In Sec.~\ref{sec:setup}, we discuss a schematic setup of a system where
a pair of chiral Majorana fermion edge modes couple to a flux qubit as a probe
of the edge states and briefly list our main findings.
In Sec.~\ref{sec:Ising}, we review the connection between the one-dimensional
critical transverse-field Ising model and Majorana fermion modes. We
identify the vortex tunneling operators between two edge states as the
disorder fields of the Ising model, and subsequently derive an effective
Hamiltonian for the flux qubit coupled to Majorana modes.
In Sec.~\ref{sec:Formalism}, we provide the necessary formalism for
evaluating the expectation values for the flux qubit state and qubit
susceptibilities.
In Sec.~\ref{sec:expectation-value} and Sec.~\ref{sec:corr-qubit}, we compute the qubit expectation values
and the two-point qubit correlation functions in the presence of the edge
state coupling, and use these results to derive the qubit susceptibility.
In Sec.~\ref{sec:higher}, we analyze higher order corrections to correlation functions of the qubit state.
We summarize our results in Sec.~\ref{sec:conclusion}.
Additionally, we provide a brief overview of the flux qubit Hamiltonian in App.~\ref{app:flux-qubit}. In App.~\ref{app:eff-two-level} we reduce the flux qubit Hamiltonian to that of a two-level system and derive the coupling between the flux qubit and the Majorana modes. In App.~\ref{app:correlation-function}, we give the form of the four point correlation function for the disorder field of the Ising model. Finally in App.~\ref{app:2nd-order-xx}, we provide the detailed derivation of the higher order corrections to the correlation functions of the qubit state.

\section{Setup of the system}
\label{sec:setup}

\begin{figure}
\includegraphics[width=0.9\linewidth]{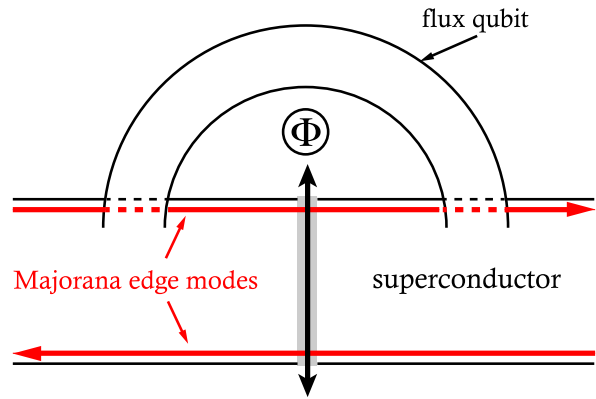}
\caption{
Schematic setup of the Majorana fermion edge modes coupled
to a flux qubit. A pair of counter-propagating edge modes appears at two opposite edges
of a topological superconductor. A flux qubit, that consists of a superconducting ring and a Josephson junction, shown as a gray rectangle, is attached to the the superconductor in such a way that it does not interrupt the edge states flow. As indicated by the arrow
across the weak link, vortices can tunnel in and out of the superconducting
ring through the Josephson junction.
}\label{fig:setup}
\end{figure}

In this work, we consider the following setup: a strip of \emph{s}-wave
superconductor is deposited on the surface of either a three-dimensional
topological insulator or a semiconductor with strong spin-orbit
coupling and broken time-reversal symmetry (or any other superconducting setup supporting Majorana edge states). As
depicted in Fig.~\ref{fig:setup}, a pair of counter-propagating
Majorana fermion edge modes appears at the two opposite edges of the
superconductor.~\cite{Akhmerov09,Fu09} To avoid mixing between
counter-propagating edge states, the width of the superconductor should be
much larger than the superconducting coherence length $\hbar v_F/\Delta$. Here
and in the following, $v_F$ denotes the Fermi velocity of the topological
insulator (semiconductor) and $\Delta$ the proximity-induced superconducting
pair-potential. In order to avoid mixing of the two counter-propagating
edge modes at the ends of the sample, we require the length of the
superconducting strip to be longer than the dephasing length.

A flux qubit, consisting of a superconducting ring with a small inductance
interrupted by a Josephson junction, is attached to the heterostructure
supporting the Majorana edge modes, as shown in Fig.~\ref{fig:setup}. By
applying an external flux $\Phi$, the two classical states of the
superconducting ring corresponding to the phase difference of $0$ and $2\pi$
across the junction can be tuned to be almost degenerate.~\cite{makhlin:01}
In this regime, the flux qubit can be viewed as a quantum two-level
system with an energy difference $\varepsilon$ (which we choose to
be positive) between the states $|0\rangle$ and $|2\pi\rangle$ and a
tunneling amplitude $\delta$ between them. As described in
App.~\ref{app:flux-qubit}, the energy difference $\varepsilon$ can be easily tuned
by the external flux $\Phi$ threaded through the ring.

The transition between the two qubit states is equivalent to the process
of a vortex tunneling through the Josephson junction in or out of the
superconducting ring. For convenience, we will refer to the Hilbert space
spanned by the qubit states $|0\rangle$ and $|2\pi \rangle$ as a spin-$1/2$
system. For example, we are going to call the Pauli matrices $\sigma^{x,y,z}$
acting on the qubit states the qubit spin.

A vortex tunneling through the weak link in the superconductor from one
edge to the other is a phase slip of $2 \pi$ of the superconducting 
phase difference at the tunneling point. Due to this event, all fermions to one side of the weak link gain a
phase of $\pi$. As will be shown below, the vortex tunneling operator can
be identified with the operator of the disorder field of a one-dimensional
critical Ising model onto which the Majorana edge modes can be mapped. 

Since vortex tunneling events couple the qubit spin to the Majorana edge modes, 
we expect various observables of the qubit to carry signatures of this coupling.
The main theory parameter that we are after is the scaling dimension 
$\Delta_{\mu}=1/8$ of the edge vortex operator (disorder field). Our main results apply to the regime when vortex tunneling is weak $\varepsilon \gg \delta$.

We find that the reduction of the spin expectation value in the $z$-direction
due to the vortex tunneling acquires a nontrivial scaling exponent
\begin{equation}
1 - \langle \sigma^{z} \rangle \propto
\frac{\delta^2}{\varepsilon^{2-2\Delta_{\mu} }} =
\frac{\delta^2}{\varepsilon^{7/4}}.
\end{equation}
Similarly, the spin expectation value along the $x$-direction
is proportional to $\varepsilon^{2 \Delta_\mu-1}=\varepsilon^{-3/4}$
thereby probing the scaling dimension of the disorder field.

The finite frequency susceptibilities that characterize
the response of the polarization of the qubit spin to a perturbation
with frequency $\omega$ provide additional information about the Majorana
edge states. The susceptibility $\chi_{zz}(\omega)$, which characterizes
the change of $\langle \sigma_z\rangle$ due to a modulation of $\sigma_z$
with frequency $\omega$, is measurable with current experimental techniques. 
It can be measured by modulating the external flux $\Phi$
 and reading out the current from a dc-SQUID coupled to the qubit.~\cite{chiorescu03,tinkham}


The frequency dependence of the
susceptibilities exhibits a non-Lorentzian resonant response around the
frequency $\omega \approx \varepsilon$ (here and in the following,
we set $\hbar =1$). It is modified by the coupling to
the Majorana edge states and shows the scaling behavior
\begin{equation}\label{eq:corr}
|\chi(\omega)|  \propto \frac{1}{
|\omega - \varepsilon|^{1- 2 \Delta_\mu}}
= \frac{1}{
|\omega - \varepsilon|^{3/4}},
\end{equation}
as long as $\varepsilon \gg |\omega - \varepsilon|$, and the distance $|\omega - \varepsilon|$ from the resonance is larger than the width of the resonance. The phase change of susceptibility at the resonance $\delta \phi=3\pi/4 $ is different from the $\pi$ phase change for a usual oscillator. The origin of the extra $\pi/4$ phase shift is the Abelian part of the statistical angle of the vortex excitations.~\cite{Nayak-RMP08}

\section{Edge states and coupling to the qubit}
\label{sec:Ising}

\subsection{Coupling of the flux qubit to the edge states}

The flux qubit has two low energy states, corresponding to a phase difference
$\phi=0$ or $\phi= 2\pi$ across the Josephson junction at $x=x_0$. The
Hamiltonian of the qubit is given by
\begin{equation}
\label{eq:qubit0}
H_Q= -\frac{\varepsilon}{2} \sigma^z - \frac{\delta}{2} e^{i\alpha} \sigma^+ - \frac{\delta}{2} e^{-i\alpha} \sigma^-. 
\end{equation}
The energy difference $\varepsilon$ can be tuned by applying an
external flux to the qubit while the tunneling amplitude $\delta>0$
can be manipulated by changing the Josephson coupling of the
junction.~\cite{makhlin:01} As discussed in App.~\ref{app:flux-qubit}, the two levels described in \eqref{eq:qubit0} represent the two lowest energy states localized at the two energy minima of a double well potential. In order for the two-level approximation to be accurate, the energies $\delta, \epsilon$ as well as the driving frequency $\omega$ have to be much smaller than the level spacing at each well. The tunneling phase $\alpha$ is proportional 
to the charge induced on the sides of the junction and its fluctuations are the main source of qubit decoherence.
For simplicity we neglect the charge noise so that we can assume that $\alpha$ is static and set it to zero without loss of generality.
The qubit Hamiltonian now reads
\begin{equation}
\label{eq:qubit}
H_Q= -\frac{\varepsilon}{2} \sigma^z - \frac{\delta}{2} \sigma^x. 
\end{equation}

When there is no phase difference across the Josephson junction ($\phi=0$),
the Hamiltonian of the chiral Majorana modes appearing at the edges of the
superconductor, as shown in Fig.~\ref{fig:setup}, reads
\begin{equation}
\label{eq:H-MF}
H_\text{MF} =  \frac{ i v^{\ }_M }{2} 
\int  \frac{dx}{2\pi} [  \psi_d(x) \partial_x  \psi_d(x)  -\psi_u(x) \partial_x \psi_u(x)],
\end{equation}
where $v_M$ is the velocity of the Majorana modes, and $\psi_u(x)$ and
$\psi_d(x)$ are the Majorana fermion fields at the upper and lower edges
of the superconductor in Fig.~\ref{fig:setup}.
The sign difference between the terms containing $\psi_{u}$ and $\psi_{d}$ is due to the fact that
the modes are counter-propagating. The Majorana fermion fields obey the
anti-commutation relations
\begin{equation}
\label{eq:commutation-relation-ud}
\begin{split}
&\{ \psi_u(x) , \psi_u(x')  \} =\{ \psi_d(x) , \psi_d(x')  \}=  
2 \pi \delta( x-x' ),
\\
&\{ \psi_u(x) , \psi_d(x')  \}=0.
\end{split}
\end{equation}

A vortex tunneling through the weak link at $x=x_0$ advances the phase
of each Cooper pair in the region $x \leq x_0$ by $2\pi$. For Majorana
fermions, just like any other fermions, this results in phase shift of
$\pi$. The effect of this phase shift is a gauge transformation
\begin{equation}
\label{eq:H-2pi}
 H_\text{MF} \mapsto P H_\text{MF} P,
\end{equation}
where the parity operator $P$ is given by
\begin{equation}
\label{eq:parity-operator}
P = \exp \Bigl[ i \pi \int^{x_0}_{-\infty} \!\!\!\! dx \, \rho_e(x) \Bigr],
\end{equation}
with the fermion density $\rho_e(x) =   \psi^{\dag}(x)
\psi(x)$ and $\psi=(\psi_{u}+i\psi_{d})/2\sqrt{\pi}$. We refer to App.~\ref{app:eff-two-level} for a derivation of the qubit Hamiltonian and the gauge transformation Eq.~\eqref{eq:H-2pi}. When the phase difference between two sides of the Josephson junction is exactly $\pi$, the Majorana modes approaching the junction are fully reflected.~\cite{Fu08} Since this phenomenon occurs only very close to the phase difference of $\pi$, where the system only spends a short amount of time during the process of a phase slip, we will neglect the effect of this backscattering. The
relation between the phase slip and the parity operator was discussed
and used in previous work focusing on the $5/2$ fractional quantum Hall
state.~\cite{Rosenow08,Rosenow09,Bena06}

Combining the Hamiltonian of the Majorana edge states (\ref{eq:H-MF}, \ref{eq:H-2pi}) with the qubit Hamiltonian \eqref{eq:qubit}, we get the full Hamiltonian of the coupled
system in the basis of $|0\rangle$ and $|2\pi\rangle$:
\begin{equation}
\label{eq:H-MF-qubit-coupled}
\mathcal{H}=
\begin{pmatrix}
H_{\text{MF}} & 0 \\
0 & P H_\text{MF} P
\end{pmatrix}
+ H_Q. 
\end{equation}
The first part of Hamiltonian represents the chiral Majorana edge states
coupled to the phase slip of the superconductor while the second part is
the bare flux qubit Hamiltonian.


Because the parity operator \eqref{eq:parity-operator} is highly nonlocal if expressed in terms
of Majorana fermions, it is desirable to map the Majorana modes on a system
where the vortex tunneling event becomes a local operator. To this end,
we establish the equivalence of the chiral Majorana edge modes with the
long wavelength limit of the one-dimensional transverse-field Ising model
at its critical point.~\cite{Zuber77,Sachdev}

\subsection{Mapping on the critical Ising model}

The lattice Hamiltonian of the Ising model at the critical point is given
by~\cite{Zuber77, Sachdev}
\begin{equation}
\label{eq:trans-Ising}
H_{I}=- J \sum_{n}  ( s_n^x s_{n+1}^x + s_n^z) ,
\end{equation}
where $s_n^{\alpha}$ are the spin-$1/2$ operators at site $n$.
With the Jordan-Wigner transformation,
\begin{equation}
\label{eq:Jordan-Wigner-trans}
\begin{split}
s^{+}_n = & c^{\ }_n \exp (i \pi \sum_{j<n} c_j^\dag c_j ) , 
\\
s^{-}_n = & c^{\dag }_n \exp (  i \pi \sum_{j<n} c_j^\dag c_j),\quad
s_n^{z}= 1- 2 c_n^{\dag } c^{\ }_n, 
\end{split}
\end{equation}
the Ising model \eqref{eq:trans-Ising} can be cast in terms of fermions as
\begin{equation}
\label{eq:H-I-complex-fermion}
H_{I}= J\sum_{n} [ (c_n^{\,}- c_n^\dag  )(c_{n+1}^{\,} + c^\dag_{n+1}) +  c_n^\dag c_n^{\,} - c_n^{\,} c_n^\dag ].
\end{equation}
Here $s^{\pm}_i \equiv (s^{x}_i \pm i s^y_i)/2$ obey the usual onsite spin commutation relations while the fermions operators $c_i^\dag$ and $c_i^{\,}$ obey canonical anti-commutation relations.

For each fermion, we introduce a pair of Majorana operators $\psi^{\
}_n=\psi_n^{\dag}$ and $\bar \psi^{\ }_n= \bar \psi_n^{\dag}$ such that
\begin{equation}
\label{eq:c-psi_12}
c_n= \frac{e^{-i \pi/4}}{2} (\psi_n + i \bar \psi_n).
\end{equation}
The Majorana fermions satisfy the Clifford algebra
\begin{align}
\{\psi_m, \psi_n \} & =\{\bar \psi_m, \bar\psi_n \} = 2 \delta_{mn},
&\{ \psi_m, \bar\psi_n \}&=0.
\end{align}
In terms of the Majorana operators, the Hamiltonian
\eqref{eq:H-I-complex-fermion} assumes the form
\begin{multline}\label{eq:Ising-majorana}
  H_{I}= -\frac{i J}{2}\sum_{n} ( \psi_n \psi_{n+1} - \bar \psi_n \bar
  \psi_{n+1} \\
  +\psi_n \bar \psi_{n+1}  - \bar \psi_n \psi_{n+1}  -2  \psi_n
\bar \psi_n  ).
\end{multline}
In the long wavelength limit, the Hamiltonian \eqref{eq:Ising-majorana}
reduces to \eqref{eq:H-MF} with the identification of the continuum
Majorana operators
\begin{equation}
\label{eq:M-fermion-lattice-conti}
 \psi_u (x) \mapsto \sqrt{ \frac{\pi}{a} } \psi_n , \quad   \psi_d (x) \mapsto  \sqrt{ \frac{\pi}{a} }  \bar{\psi}_n , \quad x \mapsto na
\end{equation}
and the velocity $v_M \mapsto 2 J a$. To complete the mapping, the
bandwidth of the Ising model should be related to the cutoff energy
$\Lambda$ of the linear dispersion of the Majorana edge states, $\Lambda
\mapsto J$. Thereby, a pair of counter-propagating Majorana edge states,
$\psi_u(x)$ and $\psi_d(x)$, can be mapped on the low energy sector of the
one-dimensional transverse-field Ising model at its critical point.

For the parity operator \eqref{eq:parity-operator}, we obtain a
representation in terms of the Ising model with the following procedure:
we first discretize $\int^{x_0}\!\! dx \, \rho_e(x)$ using the mapping
\eqref{eq:M-fermion-lattice-conti} and identify $x_0\equiv n_0 a$ as a
lattice point on the Ising model. Thereafter, we obtain an expression for
the vortex tunneling operator $P$ in terms of the Ising model
\begin{equation}
\label{eq:Parity-maps-mu}
P \mapsto
\exp\Bigl( i \pi \sum_{j \leq n_0} c^\dag_j c^{\ }_j \Bigr)=
\prod_{j \leq n_0}s_j^{z}
\equiv \mu^{x}_{n_0+1/2} ,
\end{equation}
by using Eq.~\eqref{eq:c-psi_12} and the Jordan-Wigner transformation \eqref{eq:Jordan-Wigner-trans}. Here, $\mu^{x}$ is the disorder field of the
Ising model, i.e., the dual field of the spin field.\cite{Zuber77, Sachdev,Fradkin78,Kogut79}
The Ising Hamiltonian has a form identical to Eq.~\eqref{eq:trans-Ising}
when expressed through $\mu$ operators,
\begin{equation}
\label{eq:H-Ising-dual}
H_{I}=- J \sum_{n} ( \mu_{n-1/2}^x \mu_{n+1/2}^x + \mu_{n+1/2}^z),
\end{equation}
with $\mu^z_{n+1/2}= s^z_n s^z_{n+1}$.~\cite{note1} We see
that the parity operator is indeed a local operator in the dual description
of the Ising model. After mapping on the Ising model Eq.~\eqref{eq:H-2pi}
becomes (here and in the following, we use the shortcut notation $\mu=
\mu^x$)
\begin{equation}
\label{eq:H-2pi-Ising}
  P H_\text{MF}P \mapsto \mu_{n_0+1/2} \, H_I \,\mu_{n_0+1/2},
\end{equation}
and the full Hamiltonian of Majorana edge states and the flux qubit
\eqref{eq:H-MF-qubit-coupled} maps onto
\begin{equation}
\label{eq:H-MF-qubit-coupled-Ising}
\mathcal{H} \mapsto \mathcal{H}_I=
\begin{pmatrix}
H_I & 0 \\
0 & \mu_{n_0+1/2}\, H_I \, \mu_{n_0+1/2}
\end{pmatrix}
+ H_Q. 
\end{equation}
Finally, an additional unitary transformation
\begin{gather}
\label{eq:gauge-transf}
\mathcal{H}_I \mapsto  V \mathcal{H}_I V^\dag,\\
V=V^{\dag}=
\begin{pmatrix}
  1 & 0 \\
  0 & \mu_{n_0+1/2}
\end{pmatrix},
\end{gather}
yields
\begin{equation}
\label{eq:Ising-qubit-trans}
\mathcal{H}_I
= H_I  -\frac{\varepsilon}{2} \tau^z - \frac{\delta}{2}  \tau^x
\mu_{n_0+1/2}.
\end{equation}
Here, $\tau^{i}$ are the Pauli matrices acting in the Hilbert space spanned
by $|0\rangle$ and $\mu_{n_0+1/2}|2\pi\rangle$. The operators of the qubit
spin can be expressed through $\tau^{x,y,z}$ as
\begin{equation}
\label{eq:relation-original-transf-qubit}
\sigma^z =\tau^z, \quad  \sigma^x= \tau^x \mu_{n_0+1/2},
\quad \sigma^y = \tau^y \mu_{n_0+1/2}. 
\end{equation}
We use the Hamiltonian in the form of Eq.~\eqref{eq:Ising-qubit-trans}
and the qubit spin operators \eqref{eq:relation-original-transf-qubit}
in the rest of the paper.

The way of identifying \emph{two} edge Majorana states with a \emph{complete}
transverse field Ising model presented above is different from the one commonly used
in preceding research. Usually, the \emph{chiral part} of the Ising model is
identified with a \emph{single} Majorana edge.\cite{Fendley07,Nilsson10}
The advantages of our method are the possibility to write a complete
Hamiltonian of the problem and simplified book-keeping, while its drawback
is the need for the right-moving edge and the left-moving edge to have
the same geometries. Overall the differences are not important and both
methods can be used interchangeably.

\section{Formalism}
\label{sec:Formalism}

To probe the universal properties of Majorana edge states, the energy scales
of the qubit should be much smaller than the cutoff scale of the Ising
model, $\varepsilon,\, \delta \ll \Lambda$. In the weak coupling limit $\varepsilon \gg \delta$, we construct a
perturbation theory in $\delta/\varepsilon$ by separating the Hamiltonian
$\mathcal{H}_I = \mathcal{H}_0 + V$ into an unperturbed part and a perturbation
\begin{equation}
\label{eq:H-perturbed}
\mathcal{H}_0 =  H^{\ }_I  - \frac{\varepsilon}{2} \tau^z, \qquad V = -\frac{\delta}{2}  \tau^x \mu . 
\end{equation}
Without loss of generality we set $\varepsilon > 0$, so that the ground
state of the unperturbed qubit is $|0\rangle$. For brevity we omit the spatial coordinate of the $\mu$ operator in the following since it is always the same in the setup that we consider.

We use the interaction picture with time-dependent operators
\begin{equation}
  \mathcal{O}(t) = e^{i \mathcal{H}_0 t} \mathcal{O} e^{-i \mathcal{H}_0 t}. 
\end{equation}
The perturbation $V(t)$ in this picture is given by
\begin{equation}
\label{eq:V-interaction-pic-SQ}
{V}(t) = - \frac{\delta}{2} {\mu}(t) 
[ \tau^{+} (t)  +  \tau^{-} (t) ],
\end{equation}
where $\tau^{\pm}(t) = e^{\mp i \varepsilon t} \tau^\pm$ are the
time-dependent raising and lowering operators. The structure of the raising
and lowering operators leads to physics similar to the Kondo and Luttinger
liquid resonant tunneling problems.~\cite{Fendley07, Kane92a,Kane92b}

In the calculation we need the real-time two-point and four-point
correlation functions of $\mu$ in the long-time limit
$\Lambda|t-t'|\gg 1$. The two-point correlation function is
\begin{equation}
\label{eq:mu-mu-corre}
 \langle \mu (t) \mu(t') \rangle =
 \frac{ e^{-i  \text{sgn}(t-t') \pi/8}} {\Lambda^{2 \Delta_\mu} | t- t'|^{2 \Delta_\mu}},
\end{equation}
where sgn$(x)$ denotes the sign of $x$, and $\Delta_\mu = 1/8$ the
scaling dimension of the $\mu$ field.~\cite{Ginsparg} The phase shift $\pi/8$ of the two-point correlator is the Abelian part of the statistical angle for the Ising anyons braiding rules.~\cite{Nayak-RMP08} Correlation functions involving combination of multiple fields can be obtained
via the underlying Ising conformal field theory or via a bosonization
scheme.~\cite{DiFrancesco,Ginsparg,Allen00} The expression for the four-point
correlation function is given in App.~\ref{app:correlation-function}
due to its length. For brevity we will measure energies in units of $\Lambda$ and times in units of $1/\Lambda$ in the following calculation and restore the dimensionality in the final result.

We are interested in observables of the flux qubit: the spin expectation
values and the spin susceptibilities. We use time-dependent perturbation theory to calculate these quantities.~\cite{Mahan} This method is straightforward because of the simple form of the perturbing Hamiltonian \eqref{eq:V-interaction-pic-SQ} in terms of raising and lowering operators.

Assuming that the system is in the unperturbed ground state at time $t_0 \to
-\infty$, the expectation value of a qubit spin operator $\sigma^{\alpha}(t)$
is expressed through the S-matrix $S(t, t')$,
\begin{gather}
\label{eq:expectation-value-S}
\langle \sigma^{\alpha}(t) \rangle =  
\langle S(t,t_0)^\dag  {\sigma}^{\alpha}(t) S(t,t_0)\rangle_0,\\
\label{eq:S-matrix-ito-V}
S(t,t') = \mathcal{T} \exp \left( -i \int_{t'}^{t} {V}(s) ds \right),\, t > t'.
\end{gather}
Here, $\mathcal{T}$ is the time-ordering operator and $\langle \cdot
\rangle_0$ is the expectation value with respect to the unperturbed ground
state. Similarly, the two-point correlation functions of the qubit spin
are given by
\begin{equation}
\label{eq:corr-function-S}
\langle \sigma^{\alpha}(t) \sigma^{\beta}(0) \rangle =  \langle
S^\dag(t,t_0) {\sigma}^{\alpha}(t) S(t,0) {\sigma}^{\beta}(0)  S(0, t_0) \rangle_0. 
\end{equation}
The perturbative calculation for both the expectation values and
correlation functions is done by expanding the $S$-matrices in ${V}$ order
by order. This procedure is equivalent to the Schwinger-Keldysh
formalism with the expansion of $S$ and $S^{\dag}$ corresponding to
insertions on the forward and backward Keldysh contour.

According to linear response theory, the susceptibility is given by
the Fourier transform of the retarded correlation function of the
qubit\cite{Mahan}
\begin{multline}
\label{eq:susceptibility-def}
\chi^{\ }_{\alpha\beta}(\omega) = i \int_{0}^{\infty}\! dt \, e^{i \omega t}
\langle [\sigma^{\alpha}(t), \sigma^{\beta}(0)   ]   \rangle_c \\
= - 2 \int_{0}^{\infty}  \!dt\, e^{i \omega t} {\rm Im} \langle
\sigma^{\alpha}(t) \sigma^{\beta}(0) \rangle_c,
\end{multline}
where $\langle \cdot \rangle_c$ denotes the cumulant,
\begin{equation}
  \langle \sigma^\alpha(t) \sigma^\beta(0)  \rangle_c
  = \langle \sigma^\alpha(t) \sigma^\beta(0)  \rangle-\langle \sigma^\alpha(t)
  \rangle \langle \sigma^\beta(0) \rangle,
\end{equation}
and we have used $\langle \sigma^{\beta}(0) \sigma^{\alpha}(t) \rangle_c=
\langle \sigma^{\alpha}(t) \sigma^{\beta}(0) \rangle_c^*$. We see that
in order to calculate the susceptibilities only the imaginary part of the
correlation functions for $t>0$ is required.

\section{Expectation values of the qubit spin}
\label{sec:expectation-value}

In this section, we calculate the expectation values of the
qubit spin due to coupling with the Majorana edge states to the lowest
non-vanishing order. Using the identity
\begin{equation}
\label{eq:relation-z-pm}
\sigma^z= 1- 2 \sigma^{-} \sigma^{+},
\end{equation}
we obtain
\begin{equation}
\langle \sigma^z \rangle - \langle \sigma^z \rangle^{(0)} 
= - 2 \langle \sigma^{-} \sigma^{+} \rangle 
= - 2 \langle \tau^{-} \tau^{+} \rangle ,
\end{equation}
since $\langle\sigma_z\rangle^{(0)}=1$.

The first non-vanishing correction in the perturbative calculation of $\langle \sigma^{-}
\sigma^{+} \rangle$ is of second order in $V$. By expanding $S$ and $S^\dag$ in Eq.~\eqref{eq:expectation-value-S}, we obtain
\begin{gather}
\label{eq:integral-Sz}
\langle \tau^{-} \tau^{+} \rangle^{(2)} 
= \int_{-\infty}^{0} \! dt_1 \, \int_{-\infty}^{0} \! dt_2 \, I^z,\\
I^{z}= \langle  {V} (t_2) \tau^{-} \tau^{+}  {V}(t_1) 
\rangle_0. \nonumber
\end{gather}
The integrand $I_z$ originates from the first order expansion of both $S$ and
$S^\dag$. The second order contributions from the same
$S$- or $S^\dag$-matrix vanish due to the structure of $V$ in the qubit spin space.

Substituting \eqref{eq:V-interaction-pic-SQ} and \eqref{eq:mu-mu-corre}
into the integrand $I^{z}$ yields
\begin{equation}
  I^z= \frac{\delta^2 e^{i \varepsilon
    (t_1 - t_2)
    - i \text{sgn}(t_2 - t_1) \pi/8}}
  { 4 |t_2 - t_1|^{2 \Delta_\mu} } .
\end{equation}
By evaluating the integral in Eq.~(\ref{eq:integral-Sz}), we find
\begin{equation}
\label{eq:sigma-z-2-order}
\langle \sigma^{z} \rangle^{(2)}  
= - 2 \langle \tau^{-} \tau^{+}\rangle^{(2)}  
= -  \frac{3 \Gamma( \tfrac{3}{4}) \delta^2  }{8
\varepsilon^{2-2 \Delta_\mu }},
\end{equation}
where $\Gamma(x)$ denotes the Gamma function.

The expectation value of $\sigma^{x}$ in the unperturbed ground state
vanishes. The first non-vanishing contribution to $\langle \sigma^{x}\rangle$
arises to first order in $\delta/\varepsilon$. Expanding $S$ and $S^\dag$
in Eq.~\eqref{eq:expectation-value-S} to the first order yields
\begin{gather}
\label{eq:expectation-sigma-x}
\langle \sigma^x \rangle^{(1)}= \int_{-\infty}^{0}\! d t_1 \, 
I^{x}, \\
I^{x} = -i \langle [ \tau^x \mu(0), {V}(t_1) ]  \rangle_0 = 
\frac{\sin(-\varepsilon t_1 + \tfrac{\pi}{8}) \delta}{|t_1|^{2 \Delta_\mu}}
\nonumber
\end{gather}
after substituting $\sigma^{x}$ from
Eq.~\eqref{eq:relation-original-transf-qubit} and employing the
two point correlator, Eq.~\eqref{eq:mu-mu-corre}. Evaluating
\eqref{eq:expectation-sigma-x}, we find
\begin{equation}
\label{eq:x-expectation-1st}
\langle \sigma^x \rangle^{(1)} = \frac{\Gamma( \tfrac{3}{4})
\delta}{\varepsilon^{1-2 \Delta_{\mu}}}. 
\end{equation}
Finally, $\langle \sigma^y \rangle=0$ to all orders in perturbation theory
since the Hamiltonian is invariant under $\sigma^y \mapsto -\sigma^y$.

\section{Correlation functions and susceptibilities of the flux qubit spin}
\label{sec:corr-qubit}

Since we are interested in the behavior of susceptibilities at frequencies
close to the resonance $\omega \approx \varepsilon$, we only need to obtain
the long-time asymptotic of the correlation functions of the qubit spin.
Using \eqref{eq:relation-original-transf-qubit} and \eqref{eq:mu-mu-corre},
we immediately obtain that 
\begin{equation}
\label{eq:sigma-xx-0th-order}
\langle  \sigma^x(t)  \sigma^x(0)  \rangle_c = \frac{e^{- i
\varepsilon t  - i \pi/8} }{ t^{2 \Delta_\mu} } ,
\end{equation}
is non-vanishing to zeroth order. This is due to the fact that flipping
the qubit spin automatically involves creation of an edge vortex, and
$\sigma^x$ is exactly the spin flip operator. In the same manner, one obtains that $\langle  \sigma^y(t)  \sigma^y(0)  \rangle_c =\langle  \sigma^x(t)  \sigma^x(0) \rangle_c$ to zeroth order.

Concentrating next on the mixed correlator, the relations
\eqref{eq:relation-original-transf-qubit} and \eqref{eq:relation-z-pm} yield
\begin{equation}
\langle  \sigma^x(t)  \sigma^z(0)  \rangle_c =  - 2 \langle  \mu(t)  \tau^x(t)
\tau^{-} (0) \tau^{+}(0) \rangle_0.
\end{equation}
The leading non-vanishing term in this correlation function is of first order
in $\delta$ and given by
\begin{equation}\label{eq:sigma-xz-1th-order}
  \langle  \sigma^x(t)  \sigma^z(0)  \rangle^{(1)}_c  = -
  \frac{\delta}{\varepsilon} \langle  \sigma^x(t)  \sigma^x(0)  \rangle_c.
\end{equation}
in the long-time limit. 

The leading order contribution to $\langle \sigma^{z}(t) \sigma^{z}(0)
\rangle_c$ can be evaluated using \eqref{eq:mu-mu-corre} with expansions of
$S$ and $S^\dag$ to second order in $\delta$. In the long-time limit,
the leading contribution of the correlation function is given by
\begin{equation}
\label{eq:sigma-zz-long-time}
\langle  \sigma^{z}(t)  \sigma^{z}(0)\rangle^{(2)}_c 
= \frac{\delta^2}{\varepsilon^{2}} \langle  \sigma^x(t)  \sigma^x(0)
\rangle_c
.
\end{equation}
Correlators containing a single $\sigma^y$ vanish because of the invariance under $\sigma^y \mapsto -\sigma^y$. We see that all the non-vanishing two-point correlation functions are the same up to overall prefactors. Therefore, we will focus on $\langle \sigma^x(t) \sigma^x(0)
\rangle_c$ in the following.

\subsection{Energy renormalization and damping}
\label{subsec:energy-shift}

The coupling of the flux qubit to the continuum Majorana edge states can be
thought of as a two-level system coupled to an environment via the interaction
\eqref{eq:V-interaction-pic-SQ}. This coupling leads to self-energy corrections $\Sigma$ for the qubit Hamiltonian
\begin{equation}
\label{eq:self-energy-def}
\mathcal{H}_0 \mapsto \mathcal{H}_0 
+
\Sigma
,
\quad
\Sigma =
\left(
\begin{array}{cc}
\Sigma_{\uparrow \uparrow} & \Sigma_{\uparrow \downarrow} \\
\Sigma_{\downarrow \uparrow} & \Sigma_{\downarrow \downarrow} 
\end{array}
\right),
\end{equation}
that effectively shifts the energy spectrum and can also induce
damping.~\cite{Tannoudji} Since we are only interested in qubit observables, we focus on the structure of $\Sigma$ for the two-level system and do not discuss the self-energy correction of the Majorana edge states.

To second order, the self-energy correction for two spin states can be written in terms of the perturbed Hamiltonian \eqref{eq:H-perturbed} as~\cite{Tannoudji}
\begin{equation}
\label{eq:self-energy-2nd-def}
\Sigma_{\alpha\beta} =  \langle \alpha;0  | V + V (E_\alpha + i 0^{+} - \mathcal{H}_0 )^{-1} V  | 0; \beta \rangle,
\end{equation} 
where $E_{\alpha}$ is the energy for the spin-$\alpha=\uparrow,\downarrow$ qubit states and $|\alpha; 0 \rangle$ indicates that the Ising model is in its ground state with spin-$\alpha$ for the qubit state. Due to the structure of the Hamiltonian \eqref{eq:H-perturbed}, the first order correction to the self-energy vanishes. Additionally, the off-diagonal self-energy corrections vanish also to second order.

By inserting a complete set $\sum_{E_I,\beta} | E_I ;\beta \rangle \langle \beta; E_I |=1$ of the Hilbert space of $\mathcal{H}_0$ with $E_I$ denoting the complete set of eigenstates with energy $E_I$ for the Ising sector, the diagonal elements of the self-energy become
\begin{equation}
\label{eq:self-energy-diag}
 \Sigma_{\alpha\alpha}= \sum_{E_I,\beta} \frac{ \langle \alpha;0  | V | E_I ;\beta \rangle \langle \beta; E_I |  V  | 0; \alpha \rangle} {E_\alpha + i 0^{+} - (E_I +E_\beta) }.
\end{equation}
Because $V= - (\delta/2) \tau_x \mu$, only terms with $\alpha\neq \beta$ give non-vanishing contributions such that
\begin{equation}
\label{eq:self-energy-diag-1}
\Sigma_{\alpha \alpha}= \frac{\delta^2}{4} \sum_{E^{\ }_I} \frac{ \langle 0  | \mu | E_I  \rangle \langle E_I |  \mu  | 0 \rangle} { \pm  \varepsilon - E_I + i 0^{+} }, 
\end{equation}
where $+$ corresponds to $\alpha = \downarrow$, and $-$ to $\alpha = \uparrow$. The diagonal elements of the self-energy in Eq.~\eqref{eq:self-energy-diag-1} can be cast to the form
\begin{equation}
\label{eq:self-energy-diag-2}
\Sigma_{\alpha \alpha} =  - i \frac{\delta^2}{4} \int_0^\infty\!dt\, e^{\pm i \varepsilon t} e^{- 0^{+} t}
\langle \mu(t) \mu(0) \rangle.
\end{equation}
To see that \eqref{eq:self-energy-diag-2} is equal to \eqref{eq:self-energy-diag-1}, we first insert a complete set of states of the Ising model, then write the time evolution of $\mu$ in the Heisenberg picture, and finally evaluate the integral.

Evaluating Eq.~\eqref{eq:self-energy-diag-2} with Eq.~\eqref{eq:mu-mu-corre} yields
\begin{equation}
\Sigma_{\uparrow \uparrow} = -\frac{\delta^2 \Gamma(\tfrac{3}{4})}{4 \varepsilon^{1- 2 \Delta_\mu}}, \quad \Sigma_{\downarrow \downarrow} = e^{-i\pi/4}\frac{\delta^2 \Gamma(\tfrac{3}{4})}{4 \varepsilon^{1- 2 \Delta_\mu}},
\end{equation}
where we have used $\varepsilon>0$. The absence of the imaginary part for $\Sigma_{\uparrow\uparrow}$ indicates that the spin-up state is stable. The self-energy thus gives an energy shift to the spin-up state while it gives an energy shift with a damping to the spin-down state, 
\begin{equation}
\label{eq:energy-shift-damping}
E_{\alpha} = \pm \frac{\varepsilon}{2} \mapsto \pm  \frac{\varepsilon}{2} + \Sigma_{\alpha \alpha}.
\end{equation}

The energy renormalization and damping \eqref{eq:energy-shift-damping}
alter the time evolution of the ground state correlation function
\begin{equation}
\label{eq:shifted-pm-correlator}
\langle  \tau^{+}(t)  \tau^{-}(0) \rangle_0
=e^{ - i \varepsilon t }   \mapsto e^{ - i (\varepsilon+ \nu)t-\gamma t /2} ,
\end{equation}
where the energy renormalization and damping $\nu - i \gamma/2 \equiv\Sigma_{\downarrow \downarrow}-\Sigma_{\uparrow \uparrow}$ are given by
\begin{equation}
\label{eq:energy-shift-damping-difference}
\nu=  
\frac{\cos^2 (\tfrac{\pi}{8}) \Gamma(\tfrac{3}{4}) \delta^2 }
{2 \varepsilon^{1- 2 \Delta_\mu} } , \qquad
\gamma=\frac{\Gamma(\tfrac{3}{4}) \delta^2 }{2 \sqrt{2} \varepsilon^{1- 2
\Delta_\mu} }.
\end{equation}
At zero temperature, this correlator is the only non-vanishing qubit
correlator that enters in the perturbative calculation. Therefore,
the effect of the self energy can be captured by replacing
\begin{equation}
\label{eq:energy-replace}
\varepsilon \mapsto \varepsilon+ \nu - \tfrac{i}{2} \gamma,
\end{equation}
in the qubit correlation functions computed in the long-time limit excluding
the self-energy correction. Using the replacement rule \eqref{eq:energy-replace}, one obtains the zero
temperature correlator
\begin{equation}
\label{eq:sigma-xx-long-time-with-decay}
\langle  \sigma^{x}(t)  \sigma^{x}(0)\rangle_c 
= \frac{ e^{-i ( \varepsilon + \nu )t -  \gamma t/2 -i \pi/8} }{t^{2 \Delta_\mu}}.
\end{equation}

The energy renormalization and the induced damping \eqref{eq:energy-shift-damping} do not arise explicitly in the lowest-order perturbation and require the resummation of the most divergent contributions to all
orders in perturbation theory. In a system where Wick's theorem applies, the resummation for the self-energy can be derived explicitly from a diagrammatic perturbation scheme.~\cite{Mahan} Because the correlation functions of multiple $\mu$'s do not obey the Wick's theorem (see App.~\ref{app:correlation-function}), the resummation procedure for our system becomes more complicated. In the long time limit, however, the most divergent contributions in all orders can be collected by using the operator product expansion for two $\mu$ fields that resembles the structure of the Wick's theorem.~\cite{DiFrancesco,Allen00}

\subsection{Finite temperature}

Besides $\gamma$, finite temperature is an alternative source of decoherence.
The finite temperature correlators of disorder fields are readily obtained
from the zero temperature correlators using a conformal transformation~\cite{Shankar90}
\begin{equation} \label{eq:zero-finite-temp} \frac{1}{t^{2
\Delta_\mu} }\mapsto \frac{(\pi k_B T)^{2 \Delta_\mu}}{[\sinh
(\pi k_B T t)]^{2 \Delta_\mu}} , 
\end{equation}
where $T$ denotes temperature and $k_B$ the Boltzmann constant. The
finite temperature correlator $\langle \sigma^{x}(t) \sigma^{x}(0)
\rangle_c $ in the long-time limit can be obtained by substituting
Eq.~\eqref{eq:zero-finite-temp} into \eqref{eq:sigma-xx-long-time-with-decay}
with the proviso $\varepsilon \gg k_B T$ such that the temperature has no
direct effect on the qubit dynamics.

\subsection{Susceptibility}\label{sec:susceptibility}

With the correlation functions derived above, we are now in the
position to evaluate susceptibilities of the qubit. We should keep in mind
that these correlators are valid only in the long-time limit and can only be
used to study the behavior of the susceptibilities close to the resonant
frequency $\omega \approx \varepsilon$.

Evaluating Eq.~\eqref{eq:susceptibility-def} with
Eq.~\eqref{eq:sigma-xx-long-time-with-decay} yields the susceptibility
at zero temperature around the resonance,
\begin{equation}
\label{eq:chi-xx-2nd-order}
\chi_{xx}(\omega)=
\frac{e^{i 3 \pi/8} \Gamma(\tfrac{3}{4}) }{[i (\varepsilon +
\nu -\omega ) + \gamma/2]^{1- 2 \Delta_\mu}},
\end{equation}
where $\nu$ and $\gamma$ are given in \eqref{eq:energy-shift-damping-difference}. Here, we note that the susceptibility \eqref{eq:chi-xx-2nd-order} shows non-Lorentzian response. This is in contrast to the conventional Lorentzian response of a two-level system weakly coupled to the environment.~\cite{Tannoudji,note2} If we
neglect $\nu$ and $\gamma$, which are of higher order in $\delta/\varepsilon$,
this susceptibility reduces to
\begin{equation}
\label{eq:chi-xx-0nd-order}
\chi_{xx}(\omega) =
\frac{ \Gamma(\tfrac{3}{4})}{ |\omega - \varepsilon|^{1 - 2 \Delta_\mu}
}
\begin{cases}
  1,  & \text{for } \omega < \varepsilon, \\
  e^{i3\pi /4}, & \text{for } \omega > \varepsilon, 
\end{cases}
\end{equation}
so it diverges and changes the phase by $3\pi/4$ at the resonant
frequency. We can attribute this phase change to the phase shift of the correlator of two disorder fields in Eq.~\eqref{eq:mu-mu-corre}.

The presence of damping $\gamma$ in Eq.~\eqref{eq:chi-xx-2nd-order}
provides a cutoff for the divergence of the response on resonance. The maximal
susceptibility is reached at $\omega= \varepsilon+ \nu$, and its value is
given by
\begin{equation}
\label{eq:chi-xx-resonant-scaling-Delta}
|\chi_{xx}(\varepsilon+\nu)| 
= \frac{2^{1-2 \Delta_\mu}\Gamma(\tfrac{3}{4})} {\gamma^{1-2 \Delta_\mu}}.
\end{equation}
Using the proportionality of the correlation functions \eqref{eq:sigma-xz-1th-order} and
\eqref{eq:sigma-zz-long-time}, one gets that $\chi_{xz}= \chi_{zx}= -
(\delta/\varepsilon) \chi_{xx}$ and $\chi_{zz}= (\delta/\varepsilon)^2
\chi_{xx}$. It is interesting to note
that when $\delta \to 0$ both $\chi_{xx}$ and $\chi_{xz}$ are divergent
while $\chi_{zz}$ vanishes at the resonance.

\begin{figure}
\includegraphics[width=0.9\linewidth]{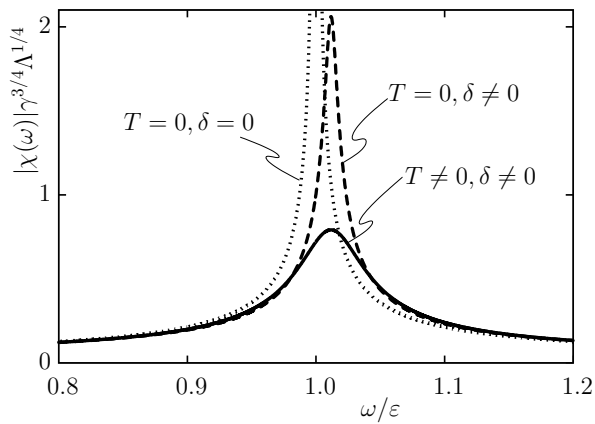}
\caption{%
Plot of the magnitude of the susceptibility $| \chi_{xx} (\omega) |$ as a function of frequency $\omega$ close to resonance $\varepsilon$. The
dotted line shows the zero temperature susceptibility in the absence
of the damping and energy renormalization while the dashed line shows
the result in the presence of the energy shift and the damping in
Eq.~\eqref{eq:energy-shift-damping}. The parameters used for the plot are
$\varepsilon=0.1 \Lambda$ and $\delta/\varepsilon=0.2$. The solid line shows
a plot of the finite temperature susceptibility with $k_B T=0.02\varepsilon$.
}\label{fig:sus-xx}
\end{figure}

In Fig.~\ref{fig:sus-xx}, the absolute value of the susceptibility
$|\chi_{xx}(\omega)|$ close to the resonance is plotted as a
function of frequency. The dotted line shows the modulus of
Eq.~\eqref{eq:chi-xx-0nd-order} for $\nu=\gamma=0$ while the dashed line
shows that of Eq.~\eqref{eq:chi-xx-2nd-order}. A renormalization of the
resonant frequency $\nu$ becomes clearly visible when comparing the peak
position of the dashed line to that of the dotted line.

The conformal dimension of the vortex excitation can be measured in
the region with $\varepsilon \gg |\omega - \varepsilon| \gtrsim \gamma$ where
\begin{equation}
\label{eq:Abs-chi-xx-scaling-frequency}
|\chi_{xx}(\omega)|  = \frac{\Gamma(\tfrac{3}{4})}{
|\omega - \varepsilon|^{1- 2 \Delta_\mu}}.
\end{equation}
Moreover, both $\chi_{xz}$ and $\chi_{zz}$ exhibit the same scaling behavior.

The finite temperature susceptibility of $\chi_{xx}(\omega,T)$
can be evaluated from the correlation function
\eqref{eq:sigma-xx-long-time-with-decay} subjected to the transformation
\eqref{eq:zero-finite-temp}. The result is plotted as the solid line in
Fig.~\ref{fig:sus-xx}. An immediate effect of the temperature is that it
also introduces a cutoff for the divergence on resonance. For instance,
the resonance peak of the susceptibility yields a different scaling
behavior with respect to the temperature
\begin{equation}
\label{eq:chi-xx-resonant-scaling-T}
|\chi_{xx}(\varepsilon+ \nu ,T)| \propto T^{-(1- 2
\Delta_\mu)},
\end{equation}
as long as $\pi k_B T \gg \gamma$. The zero temperature scaling
behavior of the resonance peak \eqref{eq:chi-xx-resonant-scaling-Delta}
will be masked by a finite temperature with a crossover at $\pi k_B
T \approx \gamma$. These scaling and crossover behaviors of the resonance
strength are features of the coupling of the Majorana edge states
and the flux qubit.~\cite{note2}

The finite temperature susceptibility shows a resonance at $\varepsilon +\nu$, as shown in Fig.~\ref{fig:sus-xx}. Around the resonance,
the frequency dependence at finite temperature will be given by the power law
\eqref{eq:Abs-chi-xx-scaling-frequency} but with the region constrained
by $\pi k_B T$ instead of $\gamma$ if $\pi k_B T>\gamma$.

\section{Higher order correlator}
\label{sec:higher}

So far, we have computed the qubit susceptibilities to their first non-vanishing orders and the lowest order self-energy correction $\varepsilon \mapsto \varepsilon + \nu - i \gamma/2$. As a consequence, we only used the two-point correlation functions $\langle \mu(t) \mu(0) \rangle$ in our evaluations. The next nontrivial corrections to the qubit correlators involve the equal position four-point correlator of the
disorder fields $\langle \mu(t_1) \mu(t_2) \mu(t_3) \mu(t_4) \rangle$. As
discussed in Appendix~\ref{app:correlation-function}, the four-point
correlator, in principle, contains information about the non-Abelian
statistics of the particles because changing the order of the fields in
the correlation function not only alters the phase but can also change the
functional form of the correlator.~\cite{Bena06} It is thus
interesting to go beyond the lowest non-vanishing order. Additionally, doing so allows to check
the consistency of the calculation of the self-energy correction done in Sec.~\ref{subsec:energy-shift}.

As an example we focus on the second order correction to the $\langle \sigma^x(t)
\sigma^x(0) \rangle_c$ correlator in the long-time limit. The 
details of the calculation are given in App.~\ref{app:2nd-order-xx} and the result in Eq.~\eqref{eq:full-2nd-xx-App}. The dominant correction is a power law divergence
\begin{equation}
  \langle  \sigma^x(t)  \sigma^x(0)  \rangle_c^{(2)} \propto
e^{- i \pi/8} \frac{e^{-i \varepsilon t}  }{ t^{2\Delta_{\mu}}} \left[ 1- ( i
\nu + \frac{\gamma}{2} ) t \right],
\end{equation}
which is just the second order in $\delta$ expansion
of the modified correlation function
\begin{equation}
\label{eq:sigma-xx-2th-order-1}
\langle  \sigma^x(t)  \sigma^x(0)  \rangle_c \propto \frac{e^{- i(
\varepsilon +\nu) t } e^{-\gamma t/2} e^{-i \pi /8 } }{
t^{2 \Delta_\mu} }.
\end{equation}
Hence, we confirm that the second order perturbative correction is
consistent with the the self-energy correction calculation. 

The leading correction to the susceptibility $\chi_{xx}$ in second order is due to the logarithmic term $\propto t^{-1/4} \log t$ in the correlator \eqref{eq:full-2nd-xx-App} and has the form
\begin{equation}
\begin{split}
  \chi_{xx}^{(2)} (\omega) =
  -\frac{\delta^2 (2 + \sqrt{2})
  \Gamma(\tfrac{7}{4}) \Gamma(\tfrac{3}{4}) e^{i3\pi/8} }{16 \varepsilon^{7/4}
  [i(\varepsilon + \nu -\omega) + \gamma/2)]^{1-2 \Delta_\mu}
  }
\\
  \times  \ln \left(\frac{ (\gamma/2)^2+ (\omega - \varepsilon-\nu)^2}{\varepsilon^2} \right)
\end{split}
\end{equation}
where we have included the self-energy correction \eqref{eq:energy-replace}, and omitted terms without logarithmic divergence. Unfortunately the effects of nontrivial exchange statistics of disorder fields are not apparent in this correction.

\section{Conclusion and discussion}
\label{sec:conclusion}

We have proposed a novel scheme to probe the edge vortex excitations of
chiral Majorana fermion edge states realized in superconducting systems
utilizing a flux qubit. To analyze the coupling we mapped the Hamiltonian 
of the Majorana edge states on the transverse-field Ising model, so that
the coupling between the qubit and the Majorana edge modes becomes a local
operator. In the weak coupling regime $\delta \ll \varepsilon$ we have found 
that the ground state expectation values of the qubit spin are given by
\begin{equation}
\langle \sigma^x \rangle=  \frac{\Gamma( \tfrac{3}{4}) \delta}{\varepsilon^{1-2 \Delta_{\mu}}\Lambda^{2\Delta_\mu}},
\, \langle \sigma^y \rangle = 0,\, 
\langle \sigma^z \rangle = 1 -\frac{3\delta}{8\varepsilon}\langle\sigma^x\rangle.
\end{equation}
Additionally, the susceptibility tensor of the qubit spin in the basis $x,y,z$ is given by
\begin{gather}
\chi(\omega) = \chi_{xx}(\omega)
\begin{pmatrix}
1 & 0 & -\delta / \varepsilon \\
0 & 1 & 0 \\
-\delta / \varepsilon & 0 & (\delta / \varepsilon)^2
\end{pmatrix} ,
\\
\chi_{xx}(\omega) = \frac{e^{i 3 \pi/8} \Gamma(\tfrac{3}{4})}{[i (\varepsilon +
\nu -\omega ) + \gamma/2]^{1- 2 \Delta_\mu} \Lambda^{2\Delta_\mu}},
\end{gather}
with the real part $\nu$ and the imaginary part $\gamma/2$ of the self-energy given by
\begin{equation}
\nu=  
\frac{\cos^2 (\tfrac{\pi}{8}) \Gamma(\frac{3}{4}) \delta^2 }
{2 \varepsilon^{1- 2 \Delta_\mu} \Lambda^{2\Delta_{\mu}} } ,
 \qquad
\gamma/2 = (\sqrt{2} -1) \nu.
\end{equation}

We see that all of these quantities acquire additional anomalous scaling 
$~(\varepsilon/\Lambda)^{2\Delta_\mu}$ due to the fact that each spin flip 
of the qubit spin couples to a disorder field $\mu$. 
Similar scaling with temperature appears in interferometric setups,\cite{Nilsson10} but
using a flux qubit allows to attribute its origin to the dynamics of vortices much more easily
and also gives additional tunability of the strength of the coupling. Another effect of the vortex tunneling being present is the phase change $\delta\phi = 3\pi/4$ of the susceptibility around the resonance.\cite{note2} This phase shift occurs due to the anomalous scaling and the presence of the Abelian statistical angle of the disorder field, in view of the fact that $\chi_{xx}$ is just a correlator of two disorder fields in the frequency domain.

The long wavelength theory which we used is only applicable when all of the energy scales are much smaller than the cutoff  energy of the Majorana modes. This is an important constraint for the flux qubit coupled to the Majorana edge states. In systems where the time-reversal symmetry is broken in the bulk (unlike for topological insulator-based proposals~\cite{note3}), the velocity of the Majorana edge states can be estimated to be $v_M \propto v_F \Delta/E_F$ and the dispersion stays approximately linear all the way up to $\Delta$. The cutoff of the Majorana modes is related to the energy scale of the Ising model $\Lambda= \Delta \mapsto J$. Equating $J= \Delta$ and $v_M = 2 J a$, we obtain the lattice constant of the Ising model $a = v_F/E_F \equiv \lambda_F$, with $\lambda_F$ the Fermi wavelength. The Fermi wavelength is typically smaller than any other length scale, and so the long wavelength approximation we have used is well-justified. For a typical flux qubit the tunneling strength $\delta$ is indeed much smaller than the superconducting gap, the level splitting $\varepsilon$ may vary from zero to quantities much larger than the superconducting gap.

Our proposal provides a way to measure properties of the non-Abelian edge vortex excitations different from the conventional detection scheme that requires fusing vortices into fermion excitations. However, none
of our results for the single flux qubit can be directly connected to the non-Abelian statistics of the quasiparticles, even after including higher-order corrections. Thus, it is of interest for future research to investigate a system where the edge vortex excitations are coupled to two qubits such that braiding of vortex excitations can be probed.~\cite{Nayak-RMP08} Another feature of systems with several qubits worth to investigate is the ability of the Majorana edge modes to mediate entanglement between different flux qubits.

\section*{Acknowledgments}

We thank C. W. J. Beenakker for useful discussions. This research was supported
by the Dutch Science Foundation NWO/FOM (C.-Y. H., A. A., and F. H.) and the Swedish Research Council (vetenskapsr{\aa}det) (J. N.).


\appendix

\section{Flux qubit} 
\label{app:flux-qubit}

The flux qubit which we consider consists of a superconducting ring interrupted
by a Josephson junction which is parameterized by its critical current
$I_\text{c}$, its capacitance $C$, and the self-inductance $L$ of the ring threaded by a magnetic
flux $\Phi$. The Hamiltonian in the phase basis reads~\cite{makhlin:01}
\begin{equation}
\label{eq:phase-basis-ham}
  H= - 4 E_C \frac{d^2}{d\phi^2} + E_J( 1- \cos \phi )  + \frac{E_L}{2} ( \phi -
  2\pi \Phi/\Phi_0)^2,
\end{equation}
where $\phi$ is the phase difference across the Josephson junction and $\Phi_0
=h/2e$ is the superconducting flux quantum. We have introduced the charging
energy $E_C=e^2/ 2 C$, the Josephson energy $E_J= \Phi_0 I_\text{c}/2\pi$, and
the inductive energy $E_L = \Phi_0^2/4\pi^2L$. 

The potential energy is given by the last two terms of the Hamiltonian \eqref{eq:phase-basis-ham}. Neglecting for a moment the inductive energy, the cosine potential favors states with $\phi= 2\pi \mathbb{Z}$. The transition between these state involves a change of the phase difference by $2\pi$ which corresponds to driving a vortex in or out of the superconducting loop. The inductive energy breaks the degeneracy of the states with a different number of vortices $n$ in the loop by favoring
states with $n\Phi_0 \approx \Phi$. When the flux $\Phi$ is tuned close to $\Phi_0/2$, the system becomes frustrated since the states $\phi=0$ and $\phi=2\pi$ are then nearly degenerate in energy. When the inductive energy is smaller than the Josephson energy but still large enough such that states with more vortices in the superconducting loop are not accessible, the potential takes the form of a double-well with the minima close to $0$ and $2\pi$. These
requirements are met when $ E_L \approx  E_J/2\pi^2$.

The charging energy $E_C$ describes the influence of quantum dynamics.
If the level spacing $\Omega = \sqrt{8 E_C E_J}$ in each well is large enough
and additionally the two wells are well separated, only the lowest energy
states $|0\rangle$ and $|2\pi\rangle$, which are localized near the classical
minima $\phi=0,2\pi$, are relevant. Hence, the low energy Hamiltonian of the
system reduces to Eq.~\eqref{eq:qubit}. For $E_L \ll E_J$, the energy detuning
of the two minima is given by $\varepsilon= 4 \pi^2 E_L (\tfrac12-\Phi/\Phi_0 )$ which can be tuned via the flux $\Phi$ in the superconducting loop. The tunneling amplitude is given by $\delta \propto \exp(- \sqrt{8 E_J/E_C})$.~\cite{makhlin:01}

Let us now discuss the experimental parameters for the flux qubit. Assuming that the superconductor order parameter $\Delta$ is about $1$ K, the
corresponding coherence length is of the order of $\xi \lesssim 1$ $\mu$m. To
avoid the mixing of the Majorana edge states, the width of Josephson junction
needs to be larger than the coherence length which is in the range of micrometers. This is consistent with most
experiments.~\cite{chiorescu03,Manucharyan09} Although the design of the flux
qubit in Ref.~\onlinecite{Manucharyan09} is more complicated than the simplest
design discussed here, the idea of a $2\pi$ phase shift for a full vortex
tunneling through the Josephson junction is the same. Thus, as a concrete
example, we quote the experimentally achieved parameters from
Ref.~\onlinecite{Manucharyan09}: $E_J\approx 9$ GHz, $E_C\approx 2.5$
GHz and $E_L \approx 0.52$ GHz. The tunneling amplitude is measured and
estimated to be $\delta \approx 369$ MHz.~\cite{Manucharyan10} Moreover, the
level spacing is estimated to be $\Omega = 13.4$ GHz.

\section{Effective two-level system}
\label{app:eff-two-level}

The Euclidean action of the superconductor phase corresponding to the Hamiltonian~\eqref{eq:phase-basis-ham} reads
\begin{equation}
\label{eq:action-phi}
S_{\phi}= \int_{-T/2}^{T/2} d \tau \left[ \frac{1}{2}\frac{1}{8 E_C} \dot{\phi}^2 +V(\phi) \right].
\end{equation}
As discussed in App.~\ref{app:flux-qubit}, the double-well potential $V(\phi)$ has two energy minima located at
$\phi=0$ and $2 \pi$ such that $V(\phi=0)=- \varepsilon/2-\Omega/2$ and
$V(\phi=2\pi)=+\varepsilon/2-\Omega/2$. For later convenience, we
have shifted the potential energy by $\tfrac12 \Omega$. In our discussion, we will assume that
the level spacing $\Omega =\sqrt{8 E_J E_C}$ is the same at both wells and that the potential profile connecting two minima can be approximated by
$V(\phi) \sim E_J(1-\cos \phi)- \tfrac12 \Omega$. The concrete form of
the potential does not affect the qualitative feature of our
discussion.~\cite{Leggett87} 

The action of the Majorana fermions can be inferred from the Hamiltonian \eqref{eq:H-MF} as
\begin{equation}
\label{eq:action-psi-ud}
S_{\psi} =\int_{-T/2}^{T/2} \frac{d \tau dx}{2 \pi} \left[ \psi_{u} \bar \partial \psi_{u} +\psi_{d} \partial \psi_{d} \right],
\end{equation}
where $\partial = (\partial_{\tau} - i \partial_{x})/2$ and $\bar \partial =
(\partial_{\tau} + i \partial_{x})/2$. The action describing the coupling between the phase field and the Majorana fermions is given by
\begin{equation}
\label{eq:action-phi-psi-coupling}
S_{\psi, \phi} = i\int_{-T/2}^{T/2} d \tau  \frac{\dot{\phi}}{2} \int_{-\infty}^{x_0} dx \rho_{e}(x, \tau),
\end{equation}
where $\rho_{e}(x, \tau)= \psi^{\dag}(x,\tau) \psi(x,\tau)$ with $\psi(x,\tau)=(\psi_{u}+i\psi_{d})/2\sqrt{\pi}$ is the fermion density of the Majorana fermions. The origin of this coupling is the electrostatic energy $VQ$ where $V=\dot{\phi}/2$ is the voltage from the Josephson relation and $Q=\int_{-\infty}^{x_0} dx \rho_{e}(x, \tau)$ is the charge of the 
superconductor island at one side of the Josephson junction.~\cite{tinkham} Here we have chosen a gauge such that the superconductor phase at the other side of the Josephson junction is fixed. Observe that the equation of motion of the phase field is not affected by the coupling term \eqref{eq:action-phi-psi-coupling} when the integration of the fermion density yields no explicit time dependence. 

The total action of the system thus becomes $S = S_{\phi}+S_{\psi}+S_{\psi, \phi} $. 
From the Euclidean (imaginary time) version of Feynman's path integral, the transition rate reads
\begin{equation}
\label{eq:Euclidean-path}
\langle\phi_f |  e^{- H T}  | \phi_i  \rangle = \mathcal{N} \int [d \psi] \int [d \phi]  e^{- S },
\end{equation}
where $H$ is the corresponding Hamiltonian, $| \phi_{i,f}\rangle$ represent the initial and final phase eigenstates, and $\mathcal{N}$ is the normalization constant. Because the leading contribution to Eq.~(\ref{eq:Euclidean-path}) at large times $T \to \infty$ comes from the lowest-lying energy eigenstates, the Hamiltonian at the left hand side can be approximated by an effective Hamiltonian that contains only a few low energy states.~\cite{Coleman} 

For the double well potential $V(\phi)$, there exist two low energy states that are localized at the two classical minima at $\phi=0$ and $2\pi$. By considering the transition rates within and between two minima
\begin{equation}
\label{eq:transition-rates}
R_{ \phi_f , \phi_i }= \langle \phi_f |  e^{- H T }  | \phi_i \rangle,
\end{equation}
for $\phi_{i,f}=0,2 \pi$, we would like to show that the effective Hamiltonian is a two-level system coupled to the Majorana fermions.

To compute transition rates for $\phi_i=\phi_f$, we first observe that the phase field is mostly localized at one of the wells and behaves as a simple harmonic oscillator. Therefore, the main contributions to the transition rates \eqref{eq:transition-rates} come from the phase field in the localized states and are given by 
\begin{equation}
\label{eq:R-non-tunneling}
R_{0,0}^{0}=\sqrt{\frac{\Omega}{2} } e^{\varepsilon T/2},\; {\rm and} \quad R_{2\pi,2\pi}^{0}=\sqrt{\frac{\Omega}{2} } e^{-\varepsilon T/2},
\end{equation}
for $\phi_i=\phi_f=0$ and $2\pi$ states, respectively.~\cite{Coleman} Notice that the phase field and Majorana fermions are effectively decoupled when the phase field is localized.

The other contributions to the transition rates come from trajectories of the phase field that contain tunneling events between two wells. These tunneling events are so called instantons and anti-instantons that occur in a very short time interval $\Delta \tau \sim 1/\Omega$. In the dilute gas approximation, each instanton or anti-instanton event centered at time $\tau_i$ contributes to the transition rates with a factor 
\begin{equation}
\label{eq:single-instanton-contribution}
K _{\pm}(\tau_i) = \frac{\delta}{2} \,  P_{\pm}(\tau_i), \quad P_{\pm}(\tau)= e^{ \pm i \pi \int_{-\infty}^{x_0} dx \rho_{e} (x, \tau)}.
\end{equation}
Here, $\delta \sim e^{- \sqrt{8 E_J/E_C}}$ is the action from the tunneling of
the phase field through the barrier, and $P_{\pm}(\tau)$ is due to the coupling
\eqref{eq:action-phi-psi-coupling} with the approximation that the time
interval of instanton $\Delta \tau$ is small such that the density field can be
replaced by $\rho_e(x,\tau_i)$. Since the integration of fermion density is an
integer, we define $P_{+}(\tau)=P_{+}^{\dag}(\tau)=P_{-}(\tau)\equiv P(\tau)$. We also note that $P^2(\tau)=1$. Moreover, we are not free to distribute the instantons and anti-instantons.
They have to be alternated in time and the first tunneling event is
determined by the initial state.~\cite{Coleman} 

Let us consider the transition rate for $R_{0,0}$. Because the phase field needs
to tunnel an even number of times in order to be back to the initial well, the
number of (anti-)instantons has to be even for a non-vanishing contribution. A
trajectory of the phase field that contains $2n$ (anti-)instantons ordered in
time, $T/2>\tau_{2n} > \tau_{2n-1}>\dots>t_1>-T/2 $, gives the contribution to
$R_{0,0}$ as 
\begin{equation}
\label{eq:R00-2n}
R_{0,0}^{2n} = 
\sqrt{\frac{\Omega}{2}}
e^{\varepsilon T} 
\left\langle \int \prod_{i=1}^{2n} d
\tau_{i} \, \tfrac{\delta}{2}
e^{ (-1)^{i+1} \varepsilon \tau_{i}} 
P(\tau_i) \right\rangle_{\psi},
\end{equation}
which is integrated over the centers of (anti-)instantons $\tau_i$. Here $\langle \cdots \rangle_\psi$ is the path integral summation over fermion fields such that 
\begin{equation}
\langle O(\psi_u,\psi_d) \rangle_{\psi}= \int [d \psi] \;  O(\psi_u,\psi_d)  \; e^{-S_{\psi} },
\end{equation}
for an arbitrary fermion field combination $O(\psi_u,\psi_d) $. The total transition rate can be written as
\begin{equation}
\label{eq:R00-summed}
R_{0,0}= \sum_{j=0}^{\infty} R_{0,0}^{2j},
\end{equation}
where $R_{0,0}^{0}$ is defined in Eq.~\eqref{eq:R-non-tunneling}. The transition rate of $R_{2\pi, 2\pi}$ can be derived in the same manner and takes the same form as $R_{0, 0}$ in Eq.~\eqref{eq:R00-summed} by summing over $R_{2\pi,2\pi}^{2j}= R_{0,0}^{2j}[\varepsilon\to \varepsilon]$.

The transition rates between two wells, $R_{2\pi,0}$ and $R_{0,2 \pi}$, can also be computed by properly
counting the (anti-)instanton events. The crucial difference is now that an odd
number of tunneling events are needed for the final state to be in a different well than
the initial state. The total transition rate of $R_{2\pi,0}$ then reads 
\begin{equation}
\label{eq:R-2pi-0-summed}
R_{2\pi,0}= \sum_{j=0}^{\infty} R_{2\pi,0}^{2j+1},
\end{equation}
by summing over contributions from trajectories with odd tunneling events
\begin{equation}
\label{eq:R-2pi-0-2n+1}
R_{2\pi,0}^{2n+1} =
\sqrt{\frac{\Omega}{2}}
\left\langle \int \prod_{i=1}^{2n +1} d
\tau_{i} \, \tfrac{\delta}{2}
e^{ (-1)^{i+1} \varepsilon \tau_{i}} 
P(\tau_i) \right\rangle_{\psi}.
\end{equation}
Finally, $R_{0, 2\pi}$ also takes the same form as $R_{2\pi,0}$ with a substitution of $R_{0, 2\pi}^{2j+1}= R_{2\pi , 0}^{2j+1} [\varepsilon \to -\varepsilon]$ in Eq.~\eqref{eq:R-2pi-0-2n+1}.

By using the interaction picture, we can explicitly show that the effective
Hamiltonian \begin{equation} \label{eq:Heff-2-level-app} H_\text{eff}=
H_{\rm MF}- \frac{\varepsilon}{2} \tau^z - \frac{\delta}{2} \tau^x P,
\end{equation} reproduces the transition rates within and between two
wells in Eqs.~(\ref{eq:R00-summed}, \ref{eq:R-2pi-0-summed}) up to an
overall constant, see Eq.~\eqref{eq:Ising-qubit-trans}. Here, $H_{\rm MF}$ is defined in Eq.~(\ref{eq:H-MF}) and
$\tau^{x,z}$ are Pauli matrix acting on the two-level basis $|\phi\rangle$,
$\phi\in\{0,2\pi\}$, of the superconducting phase difference $\phi$.~\cite{note4} Note
that $\tau_x$ enters in the Hamiltonian \eqref{eq:Heff-2-level-app}
together with $P$. This is a consequence of gauge invariance --- whenever
the superconducting phase difference changes by $2\pi$ the phase of the
Majorana to the left of the junction has to be changed by $\pi$. Every
physical observable has to be gauge invariant that is why $\tau_x$ (and
$\tau_y$ for that matter) always have to occur together with $P$. In this 
spirit, we define the (observable) qubit degrees of freedom as
\begin{equation}
\label{eq:relation-original-transf-qubit-app}
\sigma^{z}=\tau^{z}, \quad \sigma^x= \tau^x P, \quad \sigma^y = \tau^y P,
\end{equation} 
see \eqref{eq:relation-original-transf-qubit}.  We thus conclude that the
two-level Hamiltonian \eqref{eq:Heff-2-level-app} together with the
identification \eqref{eq:relation-original-transf-qubit-app} gives the effective low energy description of the system.

\section{Correlation functions of disorder fields}
\label{app:correlation-function}

The one-dimensional critical transverse-field Ising model is a conformal field theory (CFT) with central charge $c=1/2$. This CFT contains the following primary fields: $\openone$, $\epsilon = i \psi \bar \psi$, $s$, and $\mu$. Here $\openone$ is the identity operator, $\epsilon$ is the energy field (a product of the right and left moving Majorana fermion fields $\psi$ and $\bar \psi$), and $s$ is the Ising spin field with its dual field $\mu$.~\cite{Ginsparg, DiFrancesco} The dual field $\mu$ is also called the disorder field and has the same scaling behavior as the Ising spin field $s$ at the critical point. On the lattice, the disorder fields $\mu$ are non-linear combinations of Ising spin fields $s$ and reside on the bonds of lattice Ising model. They are hence not independent of the Ising spin field $s$.

In the continuum and in imaginary time, the two-point correlation function of disorder fields $\mu$ can be obtained from CFT~\cite{DiFrancesco}
\begin{equation}
\langle \mu(z_1,\bar z_1) \mu(z_2,\bar z_2) \rangle =
\frac{1}{\left[(z_1-z_2)(\bar z_1- \bar z_2) \right]^{\Delta_\mu}},
\end{equation}
with $z_{i}=\tau_i + i x_i$ and $\bar z_i =\tau_i - i x_i$.

Following Ref.~\onlinecite{Bena06}, the real-time correlators can be obtained
by analytical continuation $\tau \to \xi + i t$. Here $\xi \to 0^+$ is introduced to ensure the correct phase counting and is important for the Abelian part of the statistics. The equal position two-point correlation function is given by
\begin{equation}
\langle \mu(t_1 ,x_0) \mu(t_2,x_0) \rangle = \frac{1}{( \xi+ i (t_1-t_2))^{2 \Delta_\mu}}.
\end{equation}
By using the identity
\begin{equation}
\label{eq:correlator-identity-app}
\lim_{\xi \to 0^+} \frac{1}{(\xi + i t )^{1/4}} = \frac{ e^{-i \, {\rm sgn}(t) \pi/8} }{|t|^{1/4}},
\end{equation}
one obtains the two-point correlation function in the form of Eq.~\eqref{eq:mu-mu-corre}.

The four-point correlation function of $\mu$'s can be obtained in a similar manner. In imaginary time, the correlation function is given by~\cite{DiFrancesco}
\begin{widetext}
\begin{equation}
\label{eq:4-mu-correlator-imaginary}
\langle \mu(z_1,\bar z_1) \mu(z_2, \bar z_2) \mu(z_3,\bar z_3) \mu(z_4,\bar z_4)  \rangle^2= \left| \frac{  z_{13} z_{24} }{  z_{12} z_{34} z_{14} z_{23} } \right|^{1/2}\left( \frac{1 + \left| \chi \right| + \left| 1 - \chi \right| }{2} \right), 
\end{equation}
where $\chi=(z_{12} z_{34} /z_{13} z_{24} )$ is the conformally invariant cross ratio, and the absolute values should be understood as $\left| z_{ij}  \right|^{\alpha}= (z_{ij} \bar{z}_{ij})^{\alpha/2} $. Because we are interested in tunneling at a single point, we can set $x_i=0$. In this limit the four-point correlation function can be evaluated to be
\begin{equation}
\label{eq:4-mu-correlator-imaginary-1}
\langle \mu(z_1) \mu(z_2) \mu(z_3) \mu(z_4)  \rangle^2 = 
\left\{
\begin{aligned}
& \left| \frac{  z_{13} z_{24} }{  z_{12} z_{34} z_{14} z_{23} } \right|^{1/2} \qquad \qquad \qquad \qquad \qquad \qquad\qquad, \; {\rm for}  \quad 0 <  \chi <1
\\
& \left| \frac{  z_{13} z_{24} }{  z_{12} z_{34} z_{14} z_{23} } \right|^{1/2} \left|  1 - \chi \right| = \left| \frac{  z_{14} z_{23} }{  z_{12} z_{34} z_{13} z_{24} } \right|^{1/2} \qquad\, , \; {\rm for}  \quad \chi <0
\\
& \left| \frac{  z_{13} z_{24} }{  z_{12} z_{34} z_{14} z_{23} } \right|^{1/2}  \left| \chi \right|  =  \left| \frac{ z_{12} z_{34} }{  z_{14} z_{23} z_{13} z_{24} } \right|^{1/2} \qquad \qquad,\; {\rm for} \quad  \chi > 1 
\end{aligned}
\right.
\end{equation}

The real-time correlation function can be obtained by first taking a square root of Eq.~\eqref{eq:4-mu-correlator-imaginary-1} followed by the analytical continuation, $\tau_i \to \xi +i t_i$,~\cite{Bena06}
\begin{equation}
\label{eq:4-mu-correlator-imaginary-to-real-time}
\begin{split}
&\langle \mu(t_1) \mu(t_2) \mu(t_3) \mu(t_4)  \rangle
\\
=& F_{12}(t_1,t_2,t_3,t_4) \left[ \theta(1324) + \theta(1423) + \theta(2413) + \theta(2314)+ \theta(3241)+\theta(3142)+ \theta(4132)+\theta(4231)\right]
\\
+& F_{13}(t_1,t_2,t_3,t_4) \left[ \theta(1234) + \theta(1432) + \theta(2143) + \theta(2341)+ \theta(3214)+\theta(3412)+ \theta(4123)+\theta(4321)\right]
\\
+& F_{14}(t_1,t_2,t_3,t_4) \left[ \theta(1243) + \theta(1342) + \theta(2134) + \theta(2431)+ \theta(3124)+\theta(3421)+ \theta(4213)+\theta(4312)\right],
\end{split}
\end{equation}
where $\theta(abcd)=1$ for $t_a>t_b>t_c>t_d$ and is otherwise zero. The corresponding functions $F_{ij}$ are given by
\begin{equation}
\begin{split}
F_{12}(t_1,t_2,t_3,t_4) = \frac{[\xi + i (t_1-t_2 ) ]^{1/4} [\xi + i (t_3-t_4 ) ]^{1/4} }{[\xi + i (t_1-t_3 ) ]^{1/4} [\xi + i (t_1-t_4 ) ]^{1/4} [\xi + i (t_2-t_3 ) ]^{1/4} [\xi + i (t_2-t_4 ) ]^{1/4}} 
, \\
F_{13}(t_1,t_2,t_3,t_4) = \frac{[\xi + i (t_1-t_3 ) ]^{1/4} [\xi + i (t_2-t_4 ) ]^{1/4} }{[\xi + i (t_1-t_2 ) ]^{1/4} [\xi + i (t_1-t_4 ) ]^{1/4} [\xi + i (t_2-t_3 ) ]^{1/4} [\xi + i (t_3-t_4 ) ]^{1/4}} 
, \\
F_{14}(t_1,t_2,t_3,t_4) = \frac{[\xi + i (t_1-t_4 ) ]^{1/4} [v + i (t_2-t_3 ) ]^{1/4} }{[\xi + i (t_1-t_2 ) ]^{1/4} [\xi + i (t_1-t_3 ) ]^{1/4} [\xi + i (t_2-t_4 ) ]^{1/4} [\xi + i (t_3-t_4 ) ]^{1/4}} .
\end{split}
\end{equation}
\end{widetext}

Here $F_{12}$, $F_{13}$, and $F_{14}$ are the three characteristic functions appearing in the fourth-order correlation functions. For an Abelian state, they usually appear in quasi-symmetric combinations and exchanging two of the times alters various phase factors, which is a characteristic of fractional statistics. For the current non-Abelian case, however, exchanging two of the times not only alters phase factors but can also change the form of the correlation function from one of the characteristic functions to another. This is a special feature of non-Abelian statistics.~\cite{Bena06}

\section{Second order correction to $\langle \sigma^{x}(t) \sigma^{x}(0)\rangle_c$}
\label{app:2nd-order-xx}

\begin{figure}
\includegraphics[angle=0,scale=0.8]{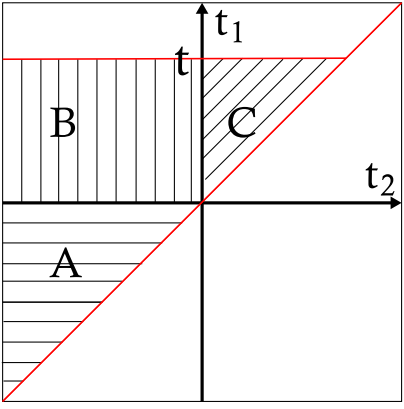}
\caption{The integral domains for regions A, B and C in the $t_1$ and $t_2$ coordinates used in Appendix~\ref{app:2nd-order-xx}.
}
\label{fig:Integral-domain}
\end{figure}

Because our ultimate goal is to compute the qubit susceptibility, we are interested in the correlator with $t > 0$ in the long-time limit $t \rightarrow\infty$. Let us first recall the perturbative part of Hamiltonian \eqref{eq:V-interaction-pic-SQ} in the interaction picture:
\begin{equation}
\label{eq:V-interaction-pic-App}
{V}(t,x_0) = -\frac{\delta}{2} {\mu}(t) [ \tau^{+}(t) +  \tau^{-} (t) ].
\end{equation}
Since the vortex tunneling in or out of the superconducting ring directly couples to the disorder field of the Ising model $ \sigma^{x}(t)=  \mu(t)  \tau^{x}(t)$ in the transformed basis, the evaluation of the second order correction for the correlator $\langle \sigma^{x}(t) \sigma^{x}(0)\rangle$ requires the knowledge of the four-point correlation function derived in Appendix~\ref{app:correlation-function}.

We expand the $S$ and $S^\dag$-matrices in \eqref{eq:corr-function-S} to second order with insertions at times $t_1$ and $t_2$. Nonzero contributions to the correlator come from three regions: (A) $t>0>t_1>t_2$, (B)
$t>t_1>0>t_2$ and (C) $t>t_1>t_2 >0$. These three regions are shown
in Fig.~\ref{fig:Integral-domain}. In what follows, we will evaluate the
second order contributions from each region in the long-time limit.

\subsection{Region A: $t>0>t_1>t_2$}

The contribution from region A is given by
\begin{equation}
\label{eq:integral-A-SQ-xx-App}
\langle \sigma^x(t)  \sigma^x(0) \rangle^{(2)}_{A}= (-i)^2 \int_{-\infty}^{0} dt_1 \int_{-\infty}^{t_1} dt_2 I_{A},
\end{equation}
with the integrand 
\begin{equation}
\begin{split}
I_{A}=&+ \langle \sigma^x(t)  \sigma^x(0)  {V}_1 {V}_2 \rangle_0 + \langle {V}_2 {V}_1 \sigma^x(t)  \sigma^x(0)   \rangle_0
\\
& - \langle {V}_2 \sigma^x(t)  \sigma^x(0)   {V}_1 \rangle_0 - \langle {V}_1 \sigma^x(t)  \sigma^x(0)   {V}_2 \rangle_0 ,
\end{split}
\end{equation}
where ${V}_{i} \equiv {V}(t_i)$ is a shorthand notation. The plus and minus signs come from the location of the insertions. The plus sign corresponds to having both insertions located on the same branch (either forward $S$ or backward $S^\dag$) while the minus sign corresponds to the situation where the two insertions are located on different branches.

Because only certain orderings of insertions of raising and lowering operators $\tau^{+}$ or $\tau^{-}$, coming both from the interaction term \eqref{eq:V-interaction-pic-App} and the $\tau^{x}$, give non-vanishing contributions, the integrand is given by
\begin{equation}
\begin{split}
\left(\frac{2}{\delta}\right)^{2} I_{A}
= &+  e^{-i \varepsilon t} e^{i \varepsilon (t_2-t_1)}  \langle  \mu(t) \mu(0) \mu(t_1) \mu(t_2)  \rangle
\\
& + e^{-i \varepsilon t} e^{i \varepsilon (t_1-t_2)}  \langle  \mu(t_2) \mu(t_1)  \mu(t) \mu(0) \rangle
\\
& -  e^{+i \varepsilon t} e^{i \varepsilon (t_1-t_2)}  \langle \mu(t_2)  \mu(t) \mu(0) \mu(t_1)   \rangle
\\
& -  e^{+i \varepsilon t} e^{i \varepsilon (t_2-t_1)}  \langle \mu(t_1)  \mu(t) \mu(0) \mu(t_2)   \rangle .
\end{split}
\end{equation}
Here, the four-point correlation function can be read off from Eq.~\eqref{eq:4-mu-correlator-imaginary-to-real-time} and simplified using the identity \eqref{eq:correlator-identity-app}. Remarkably, these correlators have the same time dependence function and differ only by phase factors. This feature is characteristic also to regions B and C. After some algebra, the integrand simplifies to
\begin{widetext}
\begin{equation}
\label{eq:Integrand-A-xx-1}
I_{A} =  2 \left(\frac{\delta}{2}\right)^{2} e^{-i \pi/8} (e^{-i \varepsilon
t} - e^{ + i \varepsilon t} )  \text{Re} \left\{ e^{i \varepsilon (t_2-t_1)}  \frac{(t-t_1)^{1/4} (-t_2)^{1/4}  e^{-i \pi/8}   }{ t^{1/4} (t-t_2)^{1/4} (-t_1)^{1/4} (t_1-t_2)^{1/4 }}   \right\}. 
\end{equation}

To evaluate the integral \eqref{eq:integral-A-SQ-xx-App}, we first simplify it by introducing new variables such that $t_1= - t T$ and $t_2= -t (T +\tau)$ with the new integrating domain $0< \tau< \infty$ and $0< T< \infty$. The second order correction from region A becomes
\begin{equation}
\label{eq:integral-A-SQ-xx-App-1}
\langle \sigma^x(t) \sigma^x(0) \rangle^{(2)}_{A}= i t^{3/2} \delta^2
e^{-i \pi/8} \sin ( \varepsilon t ) 
\text{Re} \left\{  e^{- i \pi/8} \int_{0}^{\infty} d \tau  \frac{ e^{- \varepsilon t (\eta + i )\tau }}{ \tau^{1/4 } }  \int_{0}^{\infty} d T  \frac{e^{- 2 \eta \varepsilon t T}  (1+ T)^{1/4} (T+ \tau)^{1/4}    }{  (1 + T+ \tau)^{1/4} T^{1/4}}   \right\},
\end{equation}
where we have introduced a regularization factor $\exp(\varepsilon \eta t_i)$, with $\eta \to 0^{+}$.

The integral in Eq.~\eqref{eq:integral-A-SQ-xx-App-1} will not generate any oscillatory dependence but is divergent when both $T$ and $\tau$ are large. It is thus convenient to separate the algebraic part of the integrand into three parts
\begin{equation}
\begin{split}
I_{A_1}= \frac{(1+ T)^{1/4}(T+ \tau)^{1/4} }{  (1 + T+ \tau)^{1/4} (T \tau)^{1/4}  } -  \frac{1}{\tau^{1/4}}  - \frac{\tau^{3/4} }{4 (T+ \tau ) (1+ T) },
\quad
I_{A_2} = \frac{1}{\tau^{1/4}},
\quad 
I_{A_3}= \frac{\tau^{3/4} }{4 (T+ \tau ) (1+ T) }.
\end{split}
\end{equation}
Combined with the exponential prefactor, the integration of $I_{A_1}$ is regular, the integral of $I_{A_2}$ diverges linearly while that of $I_{A_3}$ diverges logarithmically.

Integrating $I_{A_2}$ with all the exponential prefactors gives
\begin{equation}
\label{eq:integrated-A2}
\int \frac{ e^{- i \pi/8} e^{- \varepsilon t (\eta + i )\tau } e^{- 2 \eta
\varepsilon t T} }{ \tau^{1/4 } } d \tau dT
=
\frac{e^{- i \pi/8} \Gamma(\tfrac{3}{4})}{2
\eta (\varepsilon  t)^{7/4} (i+ \eta)^{3/4}} 
\propto
 \frac{1}{(\varepsilon
t)^{7/4}} [ -i \frac{\Gamma(\tfrac{3}{4})}{2 \eta} +\frac{3
\Gamma(\tfrac{3}{4})}{8} + \mathcal{O}(\eta) ], \quad \eta \to 0^+
\end{equation}
Since the the linear long time divergence is purely imaginary, it does not contribute to the correlation function.

In the long-time limit, the integrals of $I_{A_1}$ and $I_{A_3}$ with all the exponential prefactors can be carried out to the lowest order in $1/(\varepsilon t)$ and are given by
\begin{gather}
\label{eq:integrated-A1}
\int e^{- i \pi/8} e^{- \varepsilon t (\eta + i )\tau } e^{- 2 \eta \varepsilon t T} I_{A_1}  d \tau dT
\sim
\frac{\Gamma(\tfrac{7}{4}) (\pi -2 (1+\log (8) ) ) } {8 (\varepsilon t )^{7/4}},\\
\label{eq:integrated-A3}
\int e^{- i \pi/8} e^{- \varepsilon t (\eta + i )\tau } e^{- 2 \eta \varepsilon t T} I_{A_3}  d \tau dT
\sim
-\frac{\Gamma (\tfrac{7}{4}) \left(3 \log (8 \varepsilon t)-(3 \pi /\sqrt{2} ) e^{-\frac{i \pi }{4}} +3 \gamma -4\right)}{12 (\varepsilon t)^{7/4}}.
\end{gather}

We now add the real parts of the three integrals \eqref{eq:integrated-A2}, \eqref{eq:integrated-A1}, and \eqref{eq:integrated-A3} and then multiply them with the prefactors in \eqref{eq:integral-A-SQ-xx-App-1}. The result is the leading long-time contribution from region A to the qubit spin correlator:
\begin{equation}
\label{eq:xx-results-region-A-App}
\begin{split}
\langle \sigma^x (t) \sigma^x(0) \rangle^{(2)}_{A} \sim & \frac{\delta^2 e^{-i \pi/8}
(e^{i \varepsilon t} - e^{ - i \varepsilon t } ) }{ 2 t^{1/4} \varepsilon
^{7/4} } \left\{ \frac{\Gamma (\tfrac{7}{4}) (7+3 \pi - 3 \gamma - 18 \log (2) - 3 \log (\varepsilon t) )}{12 } \right\} .
\end{split}
\end{equation}
In the long-time limit, the leading contribution is given by the term $\propto t^{-1/4}\log (\varepsilon t )$.
\end{widetext}

\subsection{Region B: $t>t_1>0>t_2$ }

The contribution from the region B is given by
\begin{equation}
\label{eq:integral-B-xx-App}
\langle \sigma^x(t)  \sigma^x(0) \rangle^{(2)}_{B}= (-i)^2 \int_{0}^{t} dt_1 \int_{-\infty}^{0} dt_2 I_{B},
\end{equation}
with the integrand 
\begin{equation}
\begin{split}
I_{B}=&+ \langle \sigma^x(t) {V}_1  \sigma^x(0) {V}_2 \rangle_0 + \langle {V}_2 {V}_1 \sigma^x(t)  \sigma^x(0)   \rangle_0
\\
& - \langle {V}_2 \sigma^x(t) {V}_1 \sigma^x(0)    \rangle_0 - \langle {V}_1 \sigma^x(t)  \sigma^x(0)   {V}_2 \rangle_0.
\end{split}
\end{equation}
After ordering the the raising and lowering operators $\tau^{+}$ or $\tau^{-}$ and using Eq.~\eqref{eq:4-mu-correlator-imaginary-to-real-time}, the integrand reads
\begin{subequations}
\label{eq:Integrand-B-xx-App}
\begin{align}
I_{B} =& e^{-i \pi/4} I_{B_1} + e^{+ i \pi/4} I_{B_2} - I_{B_1}^* - I_{B_2}^*,
\end{align}
where the two integrand functions are given by
\begin{align}
\label{eq:Integrand-B1-xx-App}
I_{B_1}=& \frac{ \delta^2 e^{-i \varepsilon t} e^{i \varepsilon (t_1+t_2)}  t^{1/4} (t_1-t_2)^{1/4}  }{ 4 (t-t_1)^{1/4} (t-t_2)^{1/4} (t_1)^{1/4} (-t_2)^{1/4 }},
\\
\label{eq:Integrand-B2-xx-App}
I_{B_2}=&\frac{ \delta^2 e^{-i \varepsilon t}  e^{i \varepsilon (t_1-t_2)}  t^{1/4} (t_1-t_2)^{1/4}  }{4  (t-t_1)^{1/4} (t-t_2)^{1/4} (t_1)^{1/4} (-t_2)^{1/4 }},
\end{align}
\end{subequations}
with $x^*$ denoting complex conjugate of $x$. Again, the four-point correlators of $\mu$'s in region B have the same functional form up to phase factors.

To evaluate the integral of $I_{B_1}$, we introduce new variables $x_1$ and $x_2$ with $t_1= t ( 1 - x_1)$ and $t_2= - t x_2$ such that
\begin{widetext}
\begin{equation}
\label{eq:integral-B1-xx-App}
B_1=\int I^{\ }_{B_1} dt_1 dt_2= \frac{\delta^2 t^{3/2}}{ 4 }  \int_{0}^{1} d x_1 \int_{0}^{\infty}  d x_2  e^{-i \varepsilon t (x_1+ x_2)}  \frac{ (1- x_1+ x_2)^{1/4}  }{ (1 - x_1)^{1/4} (1+ x_2)^{1/4} (x_1)^{1/4} (x_2)^{1/4 }} .
\end{equation}
We can then split the integral $B_1$ into an oscillatory contribution $B_1^O$ and a non-oscillatory one $B_1^{NO}$.

Since the non-oscillatory contribution from \eqref{eq:integral-B1-xx-App} is dominated by $x_1 \sim x_2 \approx 0$, we can expand the integrand around this point to get the leading contribution. Because we are interested in the correlator in the long-time limit, we then deform the integration contour in the complex plane such that both $x_1$ and $x_2$ change from $0$ to $-i \infty$. The leading non-oscillatory contribution is given by
\begin{equation}
\label{eq:B1-NO}
B_1^{NO} \sim \frac{\delta^2 t^{3/2}}{ 4 }  \int_{0}^{-i \infty} d x_1 \int_{0}^{-i \infty}  d x_2  e^{-i \varepsilon t (x_1+ x_2)} \left( \frac{1}{(x_1 x_2)^{1/4}} + \frac{(x_1 x_2)^{3/4}}{4}  \right)
=
\frac{\delta^2 \Gamma(\tfrac{3}{4})^2 e^{i\pi/4} }{ 4 \varepsilon^{3/2} }  \left( - 1 + \frac{ 9 }{64 \varepsilon^2 t^{2} } \right).
\end{equation}

The oscillatory contribution $B_1^O$ is dominated by $x_1\approx 1$ and $x_2 \approx 0$, we can thus expand the integrand around this point to get the leading contribution. Again we are interested in the correlator in the long-time limit and thus deform the integration contour such that $x_1$ varies from $1 - i \infty$ to $1$ and $x_2$ varies from $0$ to $-i \infty$. After these transformations $B_1^{O}$ evaluates to
\begin{equation}
\label{eq:B1-O}
B_1^{O} \sim \frac{\delta^2 t^{3/2} e^{-i \varepsilon t }}{4} \int_{-i \infty}^{0} d u_1 \int_{0}^{-i \infty}  d x_2  e^{-i \varepsilon t (u_1+ x_2)} \frac{(x_2 -u_1)^{1/4} (-u_1)^{1/4}}{u_1 x_2^{1/4}}
=
\frac{\delta^2 e^{-i \varepsilon t }}{4 \varepsilon^{7/4}} \left( \frac{\cos
(\frac{\pi }{8}) \Gamma (\tfrac{5}{8}) \Gamma(\tfrac{3}{4}) \Gamma
(\tfrac{7}{4})}{\sqrt{2}  t^{1/4} \Gamma(\tfrac{11}{8})} \right),
\end{equation}
where $u_1=x_1-1$.

Summing up, the leading contributions to $B_1$ are
\begin{equation}
\label{eq:B1-leading}
\begin{split}
B_1= \frac{\delta^2 }{ 4 \varepsilon^{3/2} }\left\{
\Gamma(\tfrac{3}{4})^2 e^{i\pi/4}  \left(  -1 + \frac{9 }{ 64 \varepsilon^2 t^{2} }  \right)
+
e^{-i \varepsilon t }  \frac{\cos (\frac{\pi }{8}) \Gamma(\tfrac{5}{8})
\Gamma(\tfrac{3}{4}) \Gamma (\tfrac{7}{4})}{\sqrt{2} \Gamma(\tfrac{11}{8}) (\varepsilon t)^{1/4} } \right\} .
\end{split}
\end{equation}
The leading non-oscillatory contribution of $B_1$ is a constant while the leading oscillatory contribution has a power law decay $\propto t^{-1/4}$. 

To integrate $I_{B_2}$, we again use the variables $t_1= t ( 1 - x_1)$ and $t_2= - t x_2$ such that
\begin{equation}
\label{eq:Integral-B2-xx-App}
\begin{split}
B_{2}= \int I_{B_2} dt_1 dt_2=\frac{ \delta^2 t^{3/2}}{4}  \int_{0}^{1} dx_1\int_{0}^{\infty} dx_2 e^{-i\varepsilon t (x_1-x_2)}  \frac{ (1-x_1+x_2)^{1/4}   }{ (x_1)^{1/4} (1+x_2)^{1/4} (1-x_1)^{1/4} (x_2)^{1/4 }} 
\end{split} .
\end{equation}
Once again, the non-oscillatory contribution is dominated by $x_1 \sim x_2 \approx 0$. We expand the algebraic part of the integrand around $x_1 = x_2 = 0$, deform the integration contour such that $x_1$ runs from $0$ to $-i \infty$ and $x_2$ from $0$ to $i \infty$, and get
\begin{equation}
B_2^{NO} \sim  \frac{\delta^2 t^{3/2}}{4}  \int_{0}^{-i \infty} d x_1 \int_{0}^{i \infty}  d x_2  e^{-i \varepsilon t (x_1- x_2)}  
\left\{ \frac{1}{x_1^{1/4} x_2^{1/4}} + \frac{x_1^{3/4} x_2^{3/4}}{4}  \right\}
= \frac{\delta^2 \Gamma(\tfrac{3}{4})^2 }{ 4 \varepsilon^{3/2} }  \left( 1 + \frac{ 9 }{64 \varepsilon^2 t^{2} } \right).
\end{equation}

To evaluate the oscillatory part $B_2^O$ of $B_2$, we expand the integrand around $x_1 = 1$ and $x_2 = 0$ for the leading contribution. The necessary deformation of the integration contour is now given by $x_1$ changing from $1 - i \infty$ to $1$ and $x_2$ from $0$ to $i \infty$. The leading oscillatory contribution from Eq.~\eqref{eq:Integral-B2-xx-App} is now given by
\begin{equation}
B_2^{O} \sim 
\frac{\delta^2 t^{3/2}}{4} e^{-i \varepsilon t } \int_{-i \infty}^{0} d u_1 \int_{0}^{i \infty}  d x_2  e^{-i \varepsilon t (u_1 - x_2)}  \left\{ - \frac{(x_2 -u_1)^{1/4} (-u_1)^{1/4}}{u_1 x_2^{1/4}} \right\}
=
 \frac{ \delta^2 \Gamma(\tfrac{3}{4})^2 }{4 \varepsilon^{7/4}}   \frac{2
 \Gamma(\tfrac{7}{4})}{\sqrt{\pi }  t^{1/4} } e^{i \frac{7 \pi }{8}  } e^{-i \varepsilon t }, 
\end{equation}
with $u_1 =x_1-1$.

The final expression for $B_2$ is
\begin{equation}
\label{eq:B2-leading}
\begin{split}
B_2= \frac{\delta^2 \Gamma(\tfrac{3}{4})^2 }{ 4 \varepsilon^{3/2} }  \left( 1 + \frac{ 9 }{64 \varepsilon^2 t^{2} }
+
\frac{2 \Gamma(\tfrac{7}{4})e^{i \frac{7 \pi }{8}  } }{\sqrt{\pi }  (\varepsilon  t)^{1/4} }  e^{-i \varepsilon t }  \right) .
\end{split}
\end{equation}
Similarly to $B_1$, the leading non-oscillatory contribution of $B_2$ is a constant, while the leading oscillatory contribution has a power law decay $\sim t^{-1/4}$. 

From Eq.~\eqref{eq:integral-B-xx-App} and \eqref{eq:Integrand-B-xx-App}, the leading contributions to the qubit spin correlation function from region B is given by
\begin{equation}
\label{eq:xx-results-region-B-App}
\begin{split}
&\langle \sigma^x(t)  \sigma^x(0) \rangle^{(2)}_{B} = - \left( e^{-i \pi/4 } B_1 + e^{i \pi/4 } B_2 -B_1^*-B_2^*  \right)
\\
=& 
\frac{\delta^2\Gamma(\tfrac{3}{4})^2 }{2 \varepsilon^{3/2} } \left(1 - \cos( \pi/4 ) 
-\frac{ 9 i \sin (\pi/4) }{64\varepsilon ^{2} t^2 } 
+
 \frac{3 \cos (\frac{\pi }{8}) \Gamma (\tfrac{5}{8}) }{8 \sqrt{2}
 \Gamma(\tfrac{11}{8}) ( \varepsilon t)^{1/4}} (e^{i \varepsilon t } - e^{- i
 (\varepsilon t +\pi/4 )}) + \frac{\Gamma(\tfrac{7}{4}) e^{ i \pi/8 } }{\sqrt{\pi }  (\varepsilon t)^{1/4}}( e^{- i \varepsilon t }- e^{i \varepsilon t }) \right).
\end{split}
\end{equation}

\end{widetext}

\subsection{Region C: $t>t_1>t_2>0$ }

The integral in region C reads
\begin{equation}
\label{eq:integral-C-xx-App}
\langle \sigma^x(t)  \sigma^x(0) \rangle^{(2)}_{C}= (-i)^2 \int_{0}^{t} dt_1 \int_{0}^{t_1} dt_2 I_{C}.
\end{equation}
We calculate the integrand $I_C$ in a similar way to regions A and B. We get
\begin{subequations}
\label{eq:Integrand-C-xx-App}
\begin{equation}
I_C= (e^{- i \pi/4} + 1) ( I_{C_{1}} - I_{C_{2} }),
\end{equation}
with the two integrand functions being
\begin{align}
\label{eq:Integrand-C1-xx-App}
 I_{C_{1}} =&  \frac{ \delta^2 e^{-i \varepsilon t} e^{ i \varepsilon (t_1-t_2)} (t - t_2)^{1/4} (t_1)^{1/4}   }{ 4 (t-t_1)^{1/4} (t)^{1/4} (t_2)^{1/4} (t_1-t_2)^{1/4 }} ,
\\
\label{eq:Integrand-C2-xx-App}
 I_{C_{2}} =&\frac{ \delta^2 e^{i \varepsilon t} e^{- i \varepsilon (t_1+t_2)} (t - t_2)^{1/4} (t_1)^{1/4}   }{ 4 (t-t_1)^{1/4} (t)^{1/4} (t_2)^{1/4} (t_1-t_2)^{1/4 }} . 
\end{align}
\end{subequations}

To integrate $I_{C_1}$, we make the variable transformation: $t_1= t(T+1/2+\tau/2)$ and $t_2= t(T+1/2-\tau/2)$. In terms of the new variables, the integral of $I_{C_1}$ reads
\begin{widetext}
\begin{equation}
\label{eq:C1}
C_1 = \frac{\delta^2 t^{3/2} e^{-i \varepsilon t} }{4} \int_{0}^{1} d\tau   \frac{e^{ i \varepsilon  t \tau}}{\tau^{1/4} } \int_{-1/2+ \tau/2}^{1/2-\tau/2} d T   \frac{ (1/2 - T+ \tau/2)^{1/4} (1/2+T+ \tau/2)^{1/4}   }{ (1/2-T-\tau/2)^{1/4} (1/2+T-\tau/2)^{1/4} }  .
\end{equation}
The integration over $T$ can be carried out exactly with the result
\begin{equation}
\label{eq:C1-1}
C_1 = \frac{\delta^2 t^{3/2} e^{-i \varepsilon t} }{4}    \frac{2 \sqrt{\pi}
\Gamma(\tfrac{3}{4})}{\Gamma(\tfrac{1}{4})} \int_{0}^{1} d\tau  \frac{e^{ i \varepsilon  t \tau}}{\tau^{1/4} } \sqrt{1-\tau ^2} {\ }_2F_1\left(-\frac{1}{4}, \frac{1}{2}; \frac{5}{4}; \left(\frac{1-\tau}{1+ \tau} \right)^2 \right),
\end{equation}
where $_2F_1(\alpha, \beta; \gamma; x)$ is the Gaussian
hypergeometric function.\cite{abramowitz}

We deform the integration contour in Eq.~\eqref{eq:C1-1} such that $\tau$ goes from $0$ to $+i \infty$ and then back from $1+i \infty$ to $1$. The leading contribution in the long-time limit is dominated by the region near the real axis. The expansion around $x=0$ leads to an oscillatory contribution while the expansion around $x=1$ leads to a non-oscillatory contribution. To the lowest few orders, the asymptotic behavior in the long-time limit is given by
\begin{equation}
\label{eq:C1-result}
\begin{split}
C_1 \sim &+ \frac{\delta^2 \Gamma(\tfrac{3}{4})^2 e^{i\pi/4 }  }{ 4 \varepsilon^{3/2} }\left(- 1 +  \frac{9}{64 \varepsilon ^{2} t^{2}}\right) 
\\
&+  \frac{ \delta^2 e^{-i \varepsilon  t} }{4 }\left\{ \frac{e^{3 i \pi/8}
\Gamma(\tfrac{3}{4}) t^{3/4} }{\varepsilon^{3/4}} +   \frac{ e^{7 i \pi/8}
\Gamma (\tfrac{7}{4}) (6 \log (\varepsilon t)-(6+3 i) \pi +6 \gamma -14+36 \log (2))}{12 \varepsilon ^{7/4} t^{1/4}} \right\}.
\end{split}
\end{equation}
The oscillatory contribution contains a power-law divergent $t^{3/4}$ term. As we discuss later, this term contributes to the shift of the resonant frequency and to the damping for the $\langle\sigma^{+}(t)\sigma^{-}(0)\rangle$ correlation function.

To integrate $I_{C_2}$, we first change the integration variables to $\tau$ and $T$ defined by $t_1= t(T+\tau/2)$ and $t_2= t(T-\tau/2)$ such that the integral separates into two parts
\begin{equation}
C_2=\frac{ \delta^2  t^{3/2} e^{i \varepsilon t }  }{4} \left\{ \int_{0}^{1/2} dT \int_{0}^{2 T} d\tau + \int_{1/2}^{1} dT  \int_{0}^{2- 2 T} d\tau  \right\} \left(e^{- i 2 \varepsilon t T}  \frac{ (1 - T+ \tau/2)^{1/4} (T+\tau/2 )^{1/4}   }{ (1-T-\tau/2)^{1/4} (T-\tau/2)^{1/4} (\tau)^{1/4 } } \right).
\end{equation}
After changing $T\to 1 - T$ in the second integral and then introducing $X=2T$, this equation simplifies to 
\begin{equation}
\label{eq:C2}
C_2=\frac{ \delta^2  t^{3/2}  }{4}  \text{Re} \left\{ e^{i \varepsilon t }    \int_{0}^{1} dX e^{- i \varepsilon t  X} \int_{0}^{X} d\tau  \frac{ (2 - X+ \tau)^{1/4} ( X + \tau )^{1/4}   }{ (2 - X - \tau)^{1/4} (X-\tau)^{1/4} (\tau)^{1/4 } } \right\}.
\end{equation}

Again, we deform the integration contour in the integral over $X$ with $X$ changing from $0$ to $- i \infty$ and then from $1- i\infty$ to $1$. Now the oscillatory contribution comes from $X\sim 0$ while the non-oscillatory one from $X\sim 1$. By expanding the integrand around these two points, we get the leading contributions:
\begin{equation}
\label{eq:C2-result}
\begin{split}
C_2 = \frac{\delta^2 \Gamma(\tfrac{3}{4})^2}{4 \varepsilon^{3/2}}  \left( 1 +
\frac{ 9}{64\varepsilon ^{2}t^{2}}  - \frac{ \left( e^{i \varepsilon t}
e^{i\pi/8}  + e^{-i \varepsilon t} e^{-i\pi/8} \right) }{\varepsilon^{1/4}
t^{1/4}}   \frac{\,
_2F_1\left(-\frac{1}{4},\frac{3}{4};\frac{3}{2};-1\right)\Gamma (\tfrac{7}{4}) }{\sqrt{\pi }} \right) .
\end{split}
\end{equation}
The Gaussian hypergeometric function evaluates to $_2F_1\left(-\frac{1}{4},\frac{3}{4};\frac{3}{2};-1\right)\approx 1.102$.

From Eqs.~\eqref{eq:integral-C-xx-App} and \eqref{eq:Integrand-C-xx-App}, we obtain contribution to the qubit correlation function from the region C:
\begin{equation}
\label{eq:xx-results-region-C-App}
\begin{split}
\langle \sigma^x(t)  \sigma^x(0) \rangle^{(2)}_{C}=&  -  (e^{- i \pi/4} + 1) (C_1-C_2)
\\
\sim&  \frac{\delta^2 \Gamma(\tfrac{3}{4})^2}{2 \varepsilon^{3/2}}  \left( 1+ \cos( \tfrac{\pi}{4} ) - \frac{ 9 i \sin( \tfrac{\pi}{4} )  }{64 \varepsilon^{2} t^{2}} \right)
+
\left( e^{- i \pi/8} \frac{e^{-i \varepsilon t}  }{ t^{1/4}} \right)
\frac{\delta^2 \Gamma(\tfrac{3}{4}) }{4 \varepsilon ^{7/4} }  (1+ e^{- i \pi/4}) ( -i \varepsilon t  )
\\
&
+  \frac{\delta^2 e^{-i \varepsilon t}   e^{- i \pi/8} (e^{- i \pi/4} + 1)}{4
\varepsilon^{7/4} t^{1/4} } \left\{ \frac{ \Gamma(\tfrac{7}{4}) (6 \log (\varepsilon t)-(6+3 i) \pi +6 \gamma -14+36 \log (2))}{12 } \right\}
\\
&+ \frac{\delta^2 \Gamma(\tfrac{3}{4})^2 (e^{- i \pi/4} + 1) \left( e^{i
\varepsilon t} e^{i\pi/8}  + e^{-i \varepsilon t} e^{-i\pi/8} \right) }{4
\varepsilon^{7/4} t^{1/4}}   \frac{\,
_2F_1\left(-\frac{1}{4},\frac{3}{4};\frac{3}{2};-1\right)  \Gamma
(\tfrac{7}{4}) }{\sqrt{\pi }} .
\end{split}
\end{equation}
\end{widetext}

\subsection{Final result for $\langle \sigma^x(t)  \sigma^x(0) \rangle^{(2)}_{c}$}

The second order correction to the correlation function $\langle \sigma^x(t)  \sigma^x(0) \rangle^{(2)}_{c}$ can be obtained by adding up the contributions from all the three regions, given by Eqs.~\eqref{eq:xx-results-region-A-App}, \eqref{eq:xx-results-region-B-App} and \eqref{eq:xx-results-region-C-App} and then subtracting $\langle \sigma^x \rangle^2$ as calculated in Eq.~\eqref{eq:x-expectation-1st}. The full expression for the correlator in the long-time limit is
\begin{widetext}
\begin{equation}
\label{eq:full-2nd-xx-App}
\begin{split}
\langle \sigma^x (t)\sigma^x (0)\rangle^{(2)}_c=& -  i  \frac{9 \delta^2
\Gamma(\tfrac{3}{4})^2}{64 \varepsilon^{3/2}}   \frac{  \sin(\tfrac{\pi}{4})  }{ \varepsilon^{2} t^{2}} 
+
e^{- i \pi/8} \frac{e^{-i \varepsilon t}  }{ t^{1/4}}\frac{\delta^2
\Gamma(\tfrac{3}{4}) }{4 \varepsilon^{7/4} }  (1+ e^{- i \pi/4}) ( -i \varepsilon t  )
\\
&+   \frac{\delta^2 \Gamma (\tfrac{7}{4}) \log (\varepsilon t) e^{- i \pi/8}  }{8 \varepsilon ^{7/4} t^{1/4} }   \left\{  (2  +e^{- i \pi/4}  )  e^{-i \varepsilon t} -  e^{ i \varepsilon t }   \right\}
\\
& - \frac{\delta^2 \Gamma(\tfrac{7}{4})  e^{-i \pi/8}  e^{ - i \varepsilon t }  }{ 4 \varepsilon ^{7/4} t^{1/4} } \left\{ (2+ e^{- i \pi/4} )  \frac{(7+3 \pi - 3 \gamma - 18 \log (2)  )}{6 }
+
(e^{i \pi/4} + e^{i \pi/2}) \frac{\pi }{4 } \right\}
\\
& +\frac{\delta^2\Gamma(\tfrac{3}{4})^2 \Gamma (\tfrac{7}{4})  e^{- i \varepsilon t } }{4 \varepsilon^{7/4} t^{1/4} } 
\left(
 \frac{2e^{ i \pi/8 } }{\sqrt{\pi } } - \frac{3 \cos (\frac{\pi }{8}) \Gamma
 (\tfrac{5}{8}) e^{- i \pi/4 }}{4 \sqrt{2}  \Gamma(\tfrac{11}{8}) \Gamma
 (\tfrac{7}{4})  } 
 +
 \frac{ _2F_1\left(-\frac{1}{4},\frac{3}{4};\frac{3}{2};-1\right) (e^{- i 3\pi/8} + e^{-i\pi/8}) }{\sqrt{\pi }}  \right)
\\
& + \frac{\delta^2 \Gamma(\tfrac{3}{4})^2  e^{i \varepsilon t}  }{ 4
\varepsilon ^{7/4} t^{1/4} } \left\{ \frac{ [7+3 \pi - 3 \gamma - 18 \log (2)
] e^{-i \pi/8} }{8 \Gamma(\tfrac{3}{4}) } 
+
\frac{3 \cos (\frac{\pi }{8}) \Gamma (\tfrac{5}{8}) }{4 \sqrt{2}
\Gamma(\tfrac{11}{8})   } \right.
\\
&\left. \qquad \qquad \qquad \qquad
- \frac{2\Gamma(\tfrac{7}{4})e^{ i \pi/8 } }{\sqrt{\pi } }  + \frac{2 \cos
(\frac{\pi }{8}) \, _2F_1\left(-\frac{1}{4},\frac{3}{4};\frac{3}{2};-1\right)
\Gamma (\tfrac{7}{4})}{\sqrt{\pi }}  \right\} .
\end{split}
\end{equation}
\end{widetext}
This result agrees well with numerical evaluation of the integral. A power law divergence $\sim t^{3/4}$ and a logarithmic contribution $\sim \log(\varepsilon t) /t^{1/4}$ dominate the long-time behavior of the correlator. However, this logarithmic contribution will be cut off either by the induced damping or by a finite temperature.

A heuristic way to see that the term diverging as $t^{3/4}$ corresponds to self-energy correction is to add it to the zeroth order correlator of $\langle \sigma^{x}(t)\sigma^{x}(0)\rangle $ given by \eqref{eq:sigma-xx-0th-order}. The sum of these two terms equals to
\begin{equation}
\label{eq:shift-1-app}
\begin{split}
&e^{- i \pi/8} \frac{e^{-i \varepsilon t}  }{ t^{1/4}} \left( 1-i \frac{\delta^2
\Gamma(\tfrac{3}{4}) }{4 \varepsilon^{3/4} }  (2\cos^2(\tfrac{\pi}{8})-i \frac{1}{\sqrt{2}})  t  \right)
\\
=&
e^{- i \pi/8} \frac{e^{-i \varepsilon t}  }{ t^{1/4}} \left( 1-i (\nu - i
\frac{\gamma}{2})  t  \right),
\end{split}
\end{equation}
with $\nu$ and $\gamma$ the same as in Eq.~\eqref{eq:energy-shift-damping}. It then becomes apparent that \eqref{eq:shift-1-app} is exactly the expansion of the renormalized correlator \eqref{eq:sigma-xx-long-time-with-decay} to the second order in $\delta$
\begin{equation}
\frac{ e^{- i \pi/8}   }{ t^{1/4}}e^{ -i (\varepsilon +\nu)t  -  \gamma t/2 }. 
\end{equation} 
We thus conclude that the explicit evaluation of the higher order correction gives a result consistent with the self-energy calculation.

\subsection{Comments on leading contributions of higher orders}

The leading contribution to the second order corrections comes from region $C$ with integration of $C_1$ when the integration variable $\tau$ is around $\tau=0$, cf. Eq.~\eqref{eq:C1}. Since $\tau =(t_1-t_2)/t$, this expansion to the zeroth order is equivalent to making an operator product expansion of $\mu(t_1) \mu(t_2)$ for $t_1\approx t_2$ in the four-point correlation function of $\mu$ operators.~\cite{DiFrancesco} In the n$^{th}$ order of perturbation theory with insertion times $t_{1},\dots, t_{n}$, we expect that the most divergent contribution arises when all the insertion times belong to the interval $[0, t]$. By ordering the times $t_1> t_2>\dots >t_n$ and using the operator product expansion for the pairs $\mu_{t_{2i-1}} \mu_{t_{2i}}$ for $i=1,\dots,n/2$, we get a perturbative structure resembling Wick's theorem. The resummation of these terms would give the contributions for the self energy which we calculated in Sec.~\ref{subsec:energy-shift}.


\begin{thebibliography}{10}
\bibitem{note1} In the present work, the Jordan-Wigner transformation \eqref{eq:Jordan-Wigner-trans} is introduced for the Ising spin fields $s^x$. If the transformation is introduced for the disorder field, as in Refs.~\onlinecite{DiFrancesco} and \onlinecite{Allen00}, one should interchange the Ising spin field and the disorder field in our discussion.
\bibitem{note2} For a Lorentzian response, the susceptibility
is given by $\chi(\omega)\propto 1/[(\varepsilon+\nu-\omega)- i \gamma/2 ]$ that
leads to a phase change of $\pi$, compared to $3 \pi/ 4$ in
Eq.~(\ref{eq:chi-xx-0nd-order}), across the resonance. Here, the deacy rate can generally be expressed as $\gamma=\gamma_0+\gamma'(T)$ with $\gamma'(T)\propto T$ for high temperature $T> \varepsilon$.~\cite{Tannoudji} In contrast to Eq.~(\ref{eq:chi-xx-resonant-scaling-Delta}) and Eq.~(\ref{eq:chi-xx-resonant-scaling-T}), the resonance peak for a Lorentzian response thus scales as $1/\gamma_0$ at zero temperature while it scales as $1/T$ at finite temperature.
\bibitem{note3}
For topological insulator-based proposal with time-reversal symmetry in the bulk, cf. Ref.~\onlinecite{Fu09}, the velocity of Majorana edge modes is further suppressed and is given by $v^{\ }_{M} \sim v^{\ }_F (\Delta/E^{\ }_F)^2$ when $E^{\ }_F \gg \Delta$. The cutoff energy for the linear dispersion is constrained to the region $\Lambda \sim \Delta^{2}/E^{\ }_F$. In this case, we still get $a = \lambda_F$ the Fermi wavelength.
\bibitem{note4}
The basis for the two-level system, $|\phi=0\rangle$ and $|\phi=2\pi\rangle$, used in the App.~\ref{app:eff-two-level} directly denotes the superconductor phase difference across the junction. This basis is in turn different from the basis $|0\rangle$ and $|2\pi\rangle$ defined in the main text, right above Eq.~(\ref{eq:H-MF-qubit-coupled}). They are however related by $|0\rangle=|\phi=0\rangle$ and $|2\pi\rangle=P |\phi=2\pi\rangle$.
\end{thebibliography}
\end{document}